\documentclass[]{aastex631}

	\usepackage{hyperref}
   \usepackage{longtable}

\begin{document}

  \title{Decoding the Future of Exoplanets: \\
            Asteroseismic Confirmation of Subgiant and Red Giant Hosts}

  \correspondingauthor{Sheng-Bang Qian}
		\email{qiansb@ynu.edu.cn}

		\author[0000-0002-8276-5634]{Wen-Xu Lin}
		\affiliation{Yunnan Observatories, Chinese Academy of Sciences, Kunming 650216, People's Republic of China}
		\affiliation{Department of Astronomy, School of Physics and Astronomy, Yunnan University, Kunming 650091, P. R. China}
		\affiliation{Key Laboratory of Astroparticle Physics of Yunnan Province, Yunnan University, Kunming 650091, P. R. China}
		\affiliation{University of Chinese Academy of Sciences, No.1 Yanqihu East Rd, Huairou District, Beijing, 101408, People's Republic of China}

		\author[0000-0002-5995-0794]{Sheng-Bang Qian}
		\affiliation{Department of Astronomy, School of Physics and Astronomy, Yunnan University, Kunming 650091, P. R. China}
		\affiliation{Key Laboratory of Astroparticle Physics of Yunnan Province, Yunnan University, Kunming 650091, P. R. China}

		\author[0000-0002-0796-7009]{Li-Ying Zhu}
		\affiliation{Yunnan Observatories, Chinese Academy of Sciences, Kunming 650216, People's Republic of China}
		\affiliation{University of Chinese Academy of Sciences, No.1 Yanqihu East Rd, Huairou District, Beijing, 101408, People's Republic of China}

      \author[0000-0001-9346-9876]{Wen-Ping Liao}
      \affiliation{Yunnan Observatories, Chinese Academy of Sciences, Kunming 650216, People's Republic of China}
      \affiliation{University of Chinese Academy of Sciences, No.1 Yanqihu East Rd, Huairou District, Beijing, 101408, People's Republic of China}

      \author[0000-0002-0285-6051]{Fu-Xing Li}
      \affiliation{Department of Astronomy, School of Physics and Astronomy, Yunnan University, Kunming 650091, P. R. China}
      \affiliation{Key Laboratory of Astroparticle Physics of Yunnan Province, Yunnan University, Kunming 650091, P. R. China}

      \author[0000-0002-5038-5952]{Xiang-Dong Shi}
      \affiliation{Yunnan Observatories, Chinese Academy of Sciences, Kunming 650216, People's Republic of China}
      \affiliation{University of Chinese Academy of Sciences, No.1 Yanqihu East Rd, Huairou District, Beijing, 101408, People's Republic of China}

      \author[0000-0003-3767-6939]{Lin-Jia Li}
      \affiliation{Yunnan Observatories, Chinese Academy of Sciences, Kunming 650216, People's Republic of China}
		\affiliation{University of Chinese Academy of Sciences, No.1 Yanqihu East Rd, Huairou District, Beijing, 101408, People's Republic of China}

      \author{Er-Gang Zhao}
      \affiliation{Yunnan Observatories, Chinese Academy of Sciences, Kunming 650216, People's Republic of China}
		\affiliation{University of Chinese Academy of Sciences, No.1 Yanqihu East Rd, Huairou District, Beijing, 101408, People's Republic of China}

  \begin{abstract}

   Asteroseismology has emerged as a powerful tool to unravel the intricate relationships between evolved stars and their planetary systems. In this study, we leverage this technique to investigate the evolutionary stages of five exoplanet host stars, each exhibiting solar-like oscillations. Building on our previous work that identified two host stars as red clump and red giant branch (RGB) stars, this study focuses on a new and broader sample. By precisely measuring asteroseismic parameters such as the period spacing of dipole gravity modes ($\Delta\Pi_{1}$), we provide definitive confirmation of these stars' evolutionary states as subgiants or RGB stars. These results are not only crucial for understanding the internal structures of evolved stars but also for predicting the eventual fate of their planetary companions, which may face engulfment as their host stars expand. This research highlights the profound role of asteroseismology in advancing our knowledge of planetary system evolution and opens new pathways for exploring how stellar evolution impacts planetary survival. Our findings set the stage for future studies on the dynamic fates of exoplanets, providing key insights into the intricate processes of stellar and planetary evolution.

  \end{abstract}


  \section{Introduction} \label{sec:intro}

   The discovery and study of exoplanets have advanced remarkably in recent years, with significant strides made in characterizing their atmospheres, orbits, and interactions with their host stars \citep{winn2015occurrence,mulders2018planet,guo2024characterization}. A crucial component in understanding the evolution and fate of exoplanetary systems is the study of their host stars, particularly those in the later stages of stellar evolution. As stars transition from the main sequence to more advanced phases such as the subgiant and red giant stages, their internal structures undergo dramatic changes that significantly affect their planetary systems \citep{campante2019tess,hon2023close,lin2024revealing}. This stellar evolution impacts planetary orbits, with some planets facing eventual engulfment as their host stars expand \citep{villaver2009orbital}.

   Asteroseismology, which probes the internal structure of stars by analyzing their oscillation frequencies, has become an indispensable tool in the study of evolved stars and their planetary companions \citep{chaplin2013asteroseismology,garcia2019asteroseismology}. By measuring key asteroseismic parameters such as the large frequency separation ($\Delta\nu$), the frequency at maximum power ($\nu_{max}$) and the period spacing of dipole gravity modes ($\Delta\Pi_{1}$), we can determine a star's mass, radius, and evolutionary state with great precision \citep{1995A&A...293...87K,beck2011kepler,mosser2012probing,stello2013asteroseismic}. For evolved stars,  particularly those undergoing hydrogen-shell burning in the red giant branch (RGB) phase or helium-core burning in the red clump (RC) phase, these parameters are critical for understanding the dynamics and future evolution of the star-planet system.

   With the deployment of an increasing number of space telescopes, particularly satellites designed for long-term monitoring of stellar brightness variations, such as NASA's Kepler \citep{borucki2010kepler} and the Transiting Exoplanet Survey Satellite \citep[TESS,][]{ricker2015transiting}, a vast amount of light curve data has been accumulated. This has significantly advanced both exoplanetary science and asteroseismology, and the intersection of these two fields has become increasingly rich. For instance, precise determination of stellar asteroseismic parameters \citep{huber2013fundamental,jiang2020tess}, calculating stellar mass and radius through asteroseismology to derive planetary parameters \citep{chaplin2013asteroseismic,van2018asteroseismic, 2024AJ....168...27L}, or uncovering the evolutionary state of the host star through asteroseismology to discover a planet that should have been engulfed but still persists \citep{hon2023close}.
   A prior study of ours exemplifies this: using asteroseismology, we revealed that one host star, HD 120084, is a red clump star in the helium-core burning phase, while another host star, HD 29399, is a red giant branch star undergoing hydrogen-shell burning \citep{lin2024revealing}. This work holds significant value for understanding the interactions between host stars and planets at different evolutionary stages.

   To explore the interactions between planets and post-evolved stars, identifying more relevant samples is essential. The late evolutionary stages of stars encompass multiple distinct phases, during which stars undergo dynamic changes of expansion or contraction. Moving beyond the traditional distinction between red giants and red clump stars through asteroseismology, our work focuses on extending the study to a broader range of evolutionary stages. In this study, we analyzed five host stars exhibiting solar-like oscillations and successfully determined their evolutionary states. Among these stars, one was identified as being in the subgiant phase, while the remaining four were in various evolutionary stages of the red giant branch. Based on model predictions, the planets in all these systems are expected to eventually be engulfed by their host stars.

  \section{Calculation of Global Asteroseismic Parameters} \label{sec:GAP}

   The Python package $lightkurve$ \citep{2018ascl.soft12013L} was used to download all available short cadence light curves of the host stars observed by $Kepler$ from the Mikulski Archive for Space Telescopes (MAST)\footnote{https://mast.stsci.edu/}. Subsequently, we applied the Lomb-Scargle method \citep{lomb1976least, scargle1982studies} to analyze the light curves and derive the power spectrum. 
   While there are a considerable number of host stars exhibiting solar-like oscillation signals, the majority have a low signal-to-noise ratio, making it impossible to confirm mixed modes in subsequent analyses (see Section \ref{sec:ES}). 
   A small subset of host stars does exhibit clear mixed modes; however, in most cases, the number of observed mixed modes is too limited to measure the g-mode period spacing. In contrast, the power spectra of the five host stars analyzed in this study (KOI-75, Kepler-643, Kepler-815, Kepler-1004, KOI-2640) not only display mixed modes with a high signal-to-noise ratio but also provide sufficient data to enable accurate measurements of the g-mode period spacing.
   For the five host stars involved in this study, the planets or planet candidates in their systems were all discovered using the transit method. Consequently, in the power spectrum, the transit signals from the exoplanets cause power excess, as shown in Figure \ref{fig:exp_exoplanet} in the appendix. Fortunately, the frequency range of the planetary signals is much lower than that of the asteroseismic signals, meaning there is no interference between the two. Therefore, in subsequent studies, the low-frequency range can be safely ignored.
   
   The \texttt{pySYD} package \citep{2022JOSS....7.3331C} was then used to estimate the large frequency separation, $\Delta\nu$, and the frequency at maximum power, $\nu_{max}$. To generate the signal-to-noise ratio (SNR) diagram, we subtracted the estimated background noise from the power spectrum, with the background noise calculated using a smoothing filter of width $log_{10}(0.01 \mu Hz)$. The SNR data were then input into the \texttt{PBjam}\footnote[2]{https://github.com/grd349/PBjam} \citep{2021AJ....161...62N} package for further analysis, enabling us to determine the frequencies of the observed radial ($\nu_{n,0}$) and quadrupolar ($\nu_{n,2}$) oscillation modes of the host stars (see Table \ref{tab:nu0,nu2} in the appendix). To ensure reliable results, it is crucial to use appropriate prior distributions for global asteroseismic parameters and effective temperature in \texttt{PBjam}. These priors help constrain the parameter space. 
   
   To obtain the effective temperature and other parameters of the host stars, we referred to the results provided by  the Large Sky Area Multi-object Fiber Spectroscopic Telescope \citep[LAMOST,][]{wang1996special,su2004active,2012RAA....12.1197C,zhao2012lamost} in Data Release 11 \footnote[3]{https://www.lamost.org/dr11/}. The atmospheric parameters for each star are listed in Table \ref{tab:paras}. However, KOI-2640 does not have LAMOST spectral data, so we used the effective temperature and its associated error from TESS Input Catalog \citep[TIC,][]{2019AJ....158..138S} v8 as a reference.
   Thus, we incorporated the global asteroseismic parameters from \texttt{pySYD} and the effective temperatures, along with reasonable error ranges, as prior distributions. 

   It is important to note that the global asteroseismic parameters obtained from \texttt{pySYD} and the posterior distributions from \texttt{PBjam} (e.g., the posterior distribution center value of $\Delta\nu$, shown on the x-axis of Figures \ref{fig:echelle} in the appendix) are not identical. Only the parameters derived from \texttt{pySYD} should be used as the observed global asteroseismic parameters in subsequent analyses for calculating stellar mass, radius, and other related properties. The oscillation modes identified by \texttt{PBjam}, however, can be used in the following steps for the analysis of dipole mixed modes.

   \begin{deluxetable*}{lccccc}
      \setlength{\tabcolsep}{8pt}
      \tablecaption{Asteroseismic Parameters of Host Stars\label{tab:asteropara}}
      \tablehead{
        \colhead{Host Name} & \colhead{$\nu_{max}$} &\colhead{$\Delta\nu$} & \colhead{Evolutionary Stage}  & \colhead{$\Delta\Pi_{1}$} & \colhead{q}\\
        \colhead{} & \colhead{($\mu Hz$)} & \colhead{($\mu Hz$)} & \colhead{} & \colhead{(s)} & \colhead{}
      }
      \startdata
      KOI-75      & 639.918$\pm 7.078$ & 38.361$\pm 0.151$ & subgiant & 305.4 & 0.23\\
      Kepler-643  & 526.383$\pm 3.713$ & 32.990$\pm 0.308$ & RGB  & 110.1 & 0.10\\
      Kepler-815  & 350.465$\pm 2.588$ & 23.494$\pm 0.194$ & RGB  & 97.2 & 0.10 \\
      Kepler-1004 & 319.328$\pm 2.708$ & 21.423$\pm 0.373$ & RGB  & 95.8 & 0.12 \\
      KOI-2640    & 76.714$\pm 0.470$ & 7.356$\pm 0.032$ & RGB  & 84.1 & 0.15 \\
      \enddata
    \end{deluxetable*}

   \section{Calculation of Stellar Parameters} \label{sec:stllr_paras}

   After determining $\nu_{max}$ and $\Delta \nu$, along with the $T_{eff}$ derived from spectroscopic data, we can estimate the stellar mass, radius, and other physical parameters using asteroseismic scaling relations \citep{1995A&A...293...87K, chaplin2013asteroseismology}. However, for red giants, these scaling relations can be affected by their evolutionary stage and metallicity. \citet{sharma2016stellar} addressed this issue by introducing correction factors into the scaling relations, 

   \begin{equation}\label{equation:1}
      (\frac{R}{R_{\odot } } ) \simeq (\frac{\nu_{max}}{f_{\nu_{max}}\nu_{max,\odot} } )(\frac{ \Delta\nu }{f_{\Delta\nu}\Delta\nu_{\odot} } )^{-2} (\frac{T_{eff}}{T_{eff,\odot} })^{0.5},
   \end{equation}
   \begin{equation}\label{equation:2}
      (\frac{M}{M_{\odot } } ) \simeq (\frac{\nu_{max}}{f_{\nu_{max}}\nu_{max,\odot} } )^{3}(\frac{\Delta\nu}{f_{\Delta\nu}\Delta\nu_{\odot} } )^{-4} (\frac{T_{eff}}{T_{eff,\odot} })^{1.5},
   \end{equation}
   \begin{equation}\label{equation:3}
      \Delta\nu \simeq f_{\Delta\nu}(\frac{\rho}{\rho_{\odot}})^{0.5},
   \end{equation}
   \begin{equation}\label{equation:4}
      \frac{\nu_{max}}{\nu_{max,\odot}} \simeq f_{\nu_{max}}\frac{g}{g_{\odot}}(\frac{T_{eff}}{T_{eff,\odot} })^{-0.5}.
   \end{equation}

   The correction factors, $f_{\nu_{max}}$ and $f_{\Delta\nu}$, are influenced by stellar mass, metallicity, evolutionary stage, and $T_{eff}$ \citep{sharma2016stellar}. The specific correction methods, available through the \texttt{Asfgrid} tool, are detailed on this website\footnote[1]{http://www.physics.usyd.edu.au/k2gap/Asfgrid/} \citep{sharma2016stellar, stello2023extension}.

   \begin{longrotatetable}
      \setlength{\tabcolsep}{5pt}
      \linespread{1.5}
      \begin{deluxetable*}{lcccccccccc}
      \tablecaption{Parameters of 5 Host Star\label{tab:paras}}
      \tablehead{
      \colhead{Host Name} & \colhead{Source} & \colhead{T$_{eff}$} & 
      \colhead{Surface Gravity} & \colhead{[Fe/H]} & \colhead{M$_{\star}$} & 
      \colhead{R$_{\star}$}  & 
      \colhead{Luminosity} & \colhead{\parbox{3.4cm}{\centering Asteroseismic \newline surface gravity}} \\ 
      \colhead{} & \colhead{} & \colhead{(K)} & \colhead{($\log_{10}{\mathrm{(cm/s^2)}} $)} & \colhead{(dex)} & \colhead{(M$_{\odot}$)} & 
      \colhead{(R$_{\odot}$)}  & \colhead{($\log_{10}{L_{\odot }} $)} &
      \colhead{($\log_{10}{\mathrm{(cm/s^2)}} $)}
      } 
      \startdata
      KOI-75 & LAMOST &  $5947\pm8$ & 3.840$\pm 0.010$ & -0.107$\pm 0.005$ & 1.3342$\pm 0.0514$ & 2.5400$\pm 0.0359$ & 0.8615$\pm 0.0126$ & 3.7545$\pm 0.0051$ \\
      Kepler-643 & LAMOST &  $4812\pm41$ & 3.747$\pm 0.057$ & 0.142$\pm 0.040$ & 1.0039$\pm 0.0466$ & 2.5615$\pm 0.0535$ & 0.5008$\pm 0.0256$ & 3.6237$\pm 0.0039$ \\
      Kepler-815 & LAMOST &  $4997\pm37$ & 3.508$\pm 0.053$ & 0.006$\pm 0.032$ & 1.2193$\pm 0.0523$ & 3.4270$\pm 0.0647$ & 0.8195$\pm 0.0228$ & 3.4553$\pm 0.0040$ \\
      Kepler-1004 & LAMOST &  $4921\pm33$ & 3.474$\pm 0.047$ & 0.107$\pm 0.028$ & 1.3036$\pm 0.0987$ & 3.7265$\pm 0.1349$ & 0.8654$\pm 0.0346$ & 3.4115$\pm 0.0043$ \\
      KOI-2640 & TICv8 &  $4891\pm122$ & $3.35^{\star}$ & $-0.303^{\star}$ & 1.2884$\pm 0.0601$ & 7.5700$\pm 0.1257$ & 1.4705$\pm 0.0551$ & 2.7908$\pm 0.0063$ \\
      \enddata
      \tablecomments{The table contains two types of surface gravity results for the stars: one derived from spectroscopic analysis, and the other calculated through asteroseismology. For KOI-2640, since there are no LAMOST observation data, the surface gravity and metallicity values from spectroscopy are sourced from the work of \citet{2005AAS...20711013L}.
      }
      \end{deluxetable*}
   \end{longrotatetable}

   \section{Confirmation of the Evolutionary Stage of the Host Stars} \label{sec:ES}

\begin{figure*}
    \includegraphics[width=0.85\linewidth]{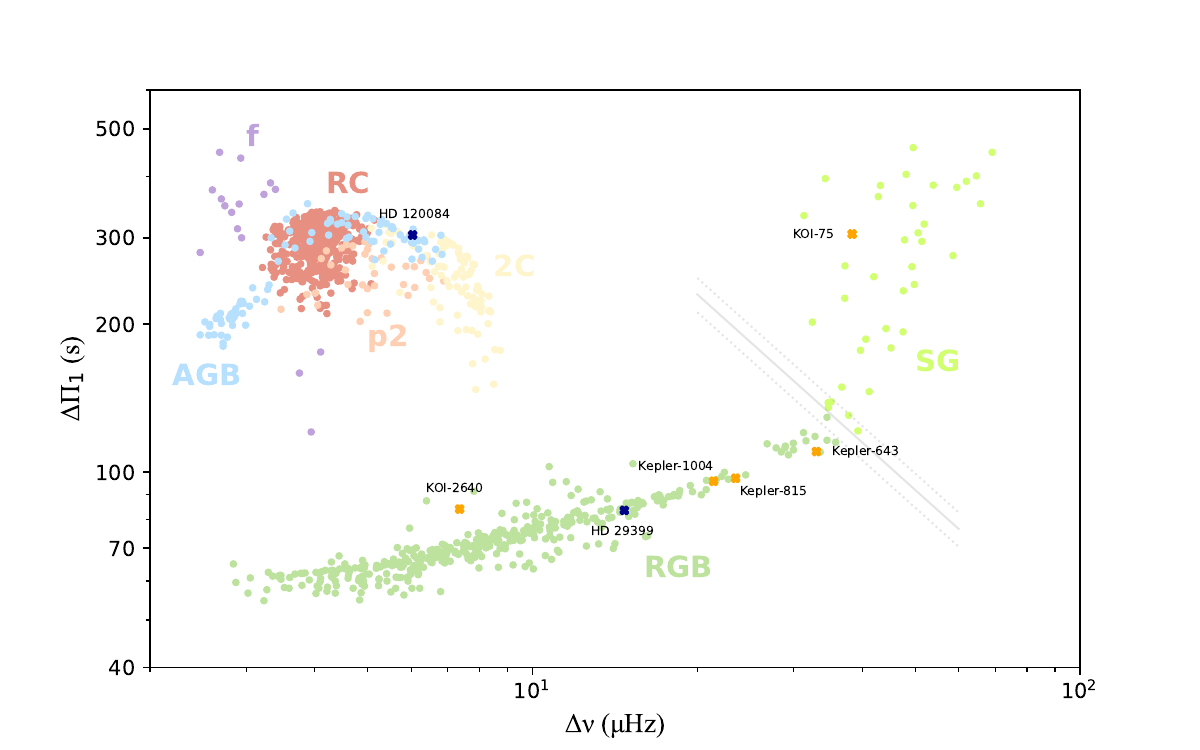}
    \centering
    \caption{$\Delta\nu - \Delta\Pi_{1}$ diagram with the 7 stars analyzed in our study and in \citet{lin2024revealing}. In this diagram, the various colored points come from the work of \citet{mosser2014mixed}. The yellow-green points (SG) represent subgiants, green points (RGB) represent red giant branch stars, purple points (f) indicate the helium subflash stars, red points (RC) represent red clump stars, light red points (p2) represent pre-secondary clump stars, and light yellow points (2C) represent secondary clump stars. The seven labeled stars include two (HD 120084, HD 29399) from our previous work \citep{lin2024revealing}, shown as dark blue “x” points. This indicates that HD 29399 is an RGB star, while HD 120084 is a secondary clump star. The five orange “x” points represent the five host stars in this study: KOI-75 is a subgiant, while the other four are RGB stars at different evolutionary stages. \label{fig:deltapi-deltanu}}
   \end{figure*}

   The calculation of the large frequency separation ($\Delta\nu$) of the star's solar-like oscillations, along with the asymptotic period spacing of dipole gravity modes ($\Delta\Pi_{1}$) \citep{mosser2012probing}, provides strong evidence for determining the star's evolutionary stage. The value of $\Delta\nu$ is proportional to the star's mean density (see Equation \ref{equation:3}), while $\Delta\Pi_{1}$ changes with the size of the radiative core and the chemical composition gradient. Therefore, on the $\Delta\nu$-$\Delta\Pi_{1}$ diagram (Figure \ref{fig:deltapi-deltanu}), stars at different evolutionary stages occupy different positions. This allows us to confirm the evolutionary stages of these host stars based on the diagram.

   In practical observations, the period spacing of consecutive mixed modes, denoted as $\Delta P$, can be derived from light curves. This period spacing is a key characteristic of dipole mixed modes. However, $\Delta P$ is not a constant value. Its relationship with the $\Delta\Pi_{1}$ is described by the following equation \citep{2015A&A...580A..96D, 2015A&A...584A..50M, vrard2016period}:

   \begin{equation}\label{equation:5}
		\Delta P = \zeta  \Delta\Pi_{1} ,
	\end{equation}

	\begin{equation}\label{equation:6}
		\zeta=\left[1+\frac{1}{q}\frac{\nu^{2} \Delta\Pi_{1}}{\Delta\nu(n_{p})}\frac{\cos^{2}\pi\frac{1}{\Delta\Pi_{1}} \left(\frac{1}{\nu}-\frac{1}{\nu_{g}}\right)}{\cos^{2}\pi\frac{\nu-\nu_{p}}{\Delta\nu(n_{p})}}   \right]^{-1} .
	\end{equation}

   where $\nu_{p}$ and $\nu_{g}$ are the asymptotic frequencies of pure pressure and gravity modes, $\Delta\nu(n_{p})$ is the frequency difference between two consecutive pure pressure radial modes with radial orders $n_{p}$ and $n_{p} +1$, and $q$ is the coupling parameter between the pressure and gravity waves \citep{1979nos..book.....U, 1989nos..book.....U}. For a detailed solution of the $\Delta\Pi_{1}$ value, please refer to the work of \citet{vrard2016period} or our previous studies \citep{lin2024revealing}. The basic principle is that without the need to identify individual mixed modes, we can directly use an appropriate $\zeta$ function to map the frequency band of the mixed modes to the period domain. Then, by performing a simple Fourier transform or calculating the power spectrum on the period-domain data, we can obtain $P(\tau)$ (see Figure \ref{fig:deltapi} in the appendix). The value of $\Delta\Pi_{1}$ corresponds to the peak of $P(\tau)$. The $\Delta\Pi_{1}$ and $q$ values for each host star are presented in Table \ref{tab:asteropara}.

   In Figure \ref{fig:deltapi-deltanu}, several gray lines can be seen separating the regions of subgiants and red giants. The solid gray line represents the empirical standard summarized by \citet{mosser2014mixed}, expressed as $(\Delta\nu / 36.5\mu Hz)(\Delta\Pi_{1} / 126s) < 1$, which is applicable to evolved stars with masses below 1.5 $M_{\odot}$. The two gray dashed lines indicate an 8\% error margin, meaning that if a star falls within the region between the two dashed lines, it cannot be definitively classified as either a subgiant or a red giant.

   \section{Planets Destined to be Engulfed by Their Host Stars} \label{sec:engulfed}

   We note that in the five exoplanetary systems studied in this work, the orbital periods of the planets are relatively short, with detailed information about each planetary system provided in Table \ref{tab:exo_para} in the appendix. The shorter orbital periods imply that, as the host star expands, and if there is no gravitational interference from other celestial bodies, these planets will inevitably be engulfed by their host stars. Next, we need to use evolutionary models to demonstrate that the planets will indeed be consumed. Our task is to calculate the moment when the stellar radius expands to reach the periapsis of the planetary orbit. In the actual evolution of planetary orbits, the expansion of the star transfers the planet's orbital angular momentum to the star's rotational angular momentum through tidal dissipation \citep{2016CeMDA.126..275B}. This process causes the planetary orbit to gradually shrink, leading to the planet being engulfed before the star's radius fully reaches the planetary orbit.

\begin{figure*}
      \includegraphics[width=0.65\linewidth]{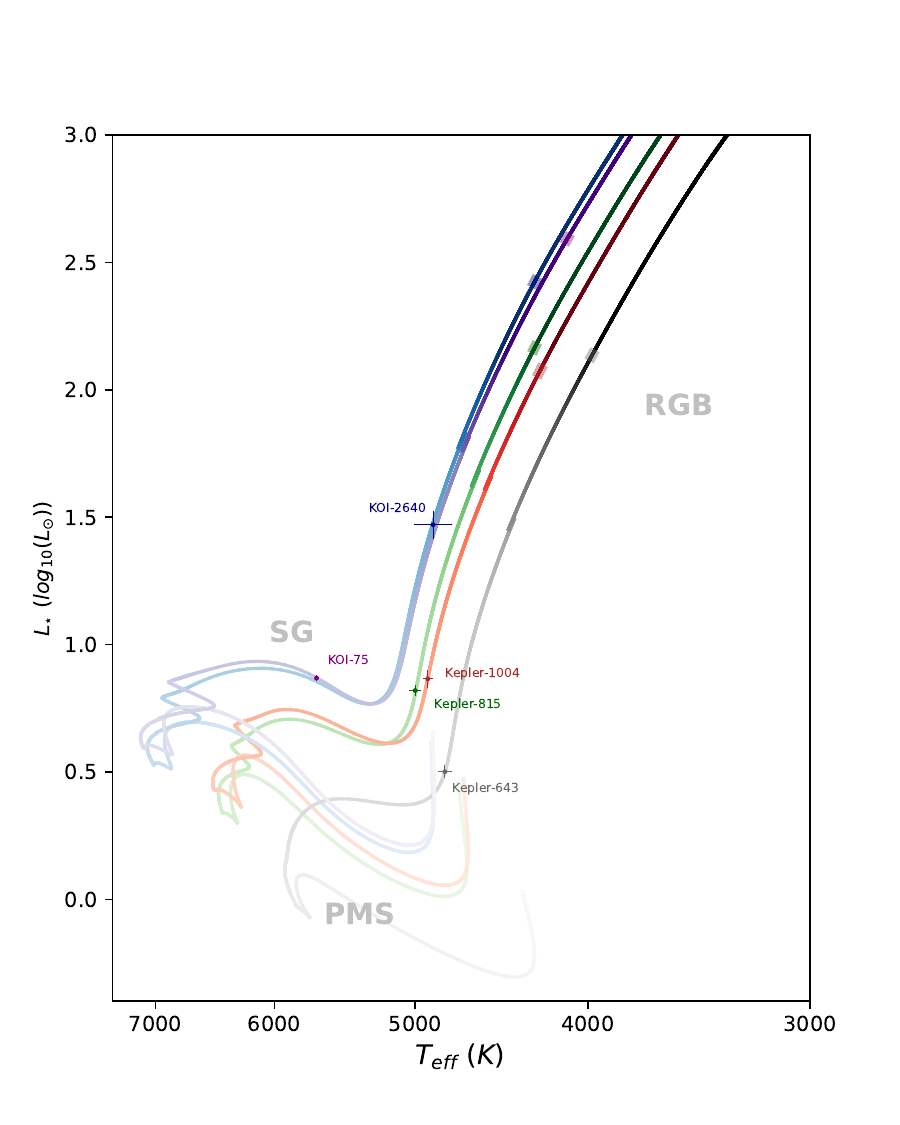}
      \centering
      \caption{The positions of the five host stars on the Hertzsprung-Russell diagram, along with the evolution tracks corresponding to the color-coded curves, are computed using MESA. During stellar evolution, stars pass through the pre-main sequence (PMS), main sequence, subgiant (SG), and red giant branch (RGB) phases. According to the evolutionary model, KOI-75 is located in the subgiant phase, while the other four host stars are in the RGB phase. It is also important to note that on the evolutionary tracks of the five host stars in the RGB phase, there are bold-colored lines marking the stage at which the star's radius expands to the same size as the periapsis distance of the planetary orbit. This indicates that before reaching this evolutionary stage, the exoplanets would have been engulfed by the expanding star, showing that the planets in all five systems are destined to be consumed.\label{fig:mesa}}
   \end{figure*}

   Next, we only need to calculate whether the star can expand to reach the periapsis of the planetary orbit. Given the host star's mass, the planet's orbital period, and the orbital eccentricity, we can derive the expression for the periapsis distance:

   \begin{equation}\label{equation:7}
		r_{peri} = (\frac{GMP^{2}}{4\pi ^{2}})^{\frac{1}{3}} (1-e) ,
	\end{equation}

   where $G$ is the gravitational constant, $M$ is the stellar mass, $P$ is the orbital period of the planet, and $e$ is the orbital eccentricity of the planet. Next, we used the \texttt{1M\_pre\_ms\_to\_wd} test suite from \texttt{MESA} (r22.11.1) (Paxton et al. \citeyear{Paxton2011,2013ApJS..208....4P,2015ApJS..220...15P,2018ApJS..234...34P,2019ApJS..243...10P}) to simulate the evolutionary tracks of each host star (as shown in Figure \ref{fig:mesa}). The condition is that the evolutionary track must fall within 0.2 times the observational error range of the star's parameters (temperature, luminosity, mass, radius, and metallicity). Additionally, we incorporate the errors in the host star's mass and the planet's orbital period into Equation \ref{equation:7} to match the \texttt{MESA} model, specifically the track where the stellar radius reaches the periapsis. In Figure \ref{fig:mesa}, we have highlighted these tracks with bold lines.

   In the MESA evolutionary model, starting from the present time, the time at which each host star expands to reach the periapsis of the planetary orbit is as follows: KOI-75 in $400.872_{-0.068}^{+0.085}$ million years, Kepler-643 in $475.889_{=0.173}^{+0.149}$ million years, Kepler-815 in $277.198_{-0.163}^{+0.160}$ million years, Kepler-1004 in $243.293_{-0.321}^{+0.367}$ million years, and KOI-2640 in $83.521_{-0.101}^{+0.109}$ million years. After this point, the host stars will continue to expand further, reaching their maximum size at the tip of the RGB phase when helium flash occurs. From this, we can conclude that the planets in these systems are all destined to be engulfed by their host stars.

   \section{Conclusion} \label{sec:conclusion}

   In this study, we have extended the application of asteroseismology to further investigate the evolutionary stages of exoplanet host stars, specifically focusing on confirming whether they are in the subgiant or red giant branch (RGB) phases. Our analysis, building upon prior work, identified five representative stars that exhibit solar-like oscillation signals, a key diagnostic in determining their evolutionary state. Through detailed asteroseismic analysis, we confirmed that these stars are indeed in the subgiant or RGB phases, thereby providing essential insight into the ongoing interaction between evolving stars and their orbiting planets.

   The confirmation of host stars in these phases is particularly significant, as stars at these stages undergo substantial changes in their internal structures and outer envelopes, which, in turn, profoundly influence their surrounding planetary systems. Planets in close orbits are especially vulnerable to tidal forces from their expanding host stars, which may lead to orbital decay, potential engulfment, or altered dynamical configurations. Our findings contribute to a broader understanding of these processes by expanding the sample of known subgiant and RGB stars hosting exoplanets, providing key data points for refining models of stellar evolution and its impact on planetary system dynamics.

   Moreover, the results highlight the utility of asteroseismology as a powerful tool for accurately determining the evolutionary stages of stars that are not easily classified through traditional observational methods. The precision of asteroseismic data enables a clearer distinction between closely related phases, such as the subgiant and early RGB phases, which can be critical when assessing the likely future of an exoplanetary system. This ability to pinpoint evolutionary stages with greater accuracy offers a more reliable framework for predicting the fate of planets orbiting stars that are approaching the final stages of stellar evolution.

   In conclusion, this study has successfully confirmed the subgiant and RGB evolutionary stages of five exoplanet host stars using asteroseismology, thereby expanding the dataset of evolved host stars and reinforcing the role of asteroseismology in stellar and planetary system research. The implications of these findings extend to a deeper understanding of how planetary systems evolve in response to the advanced stages of stellar life cycles. Future work will focus on increasing the sample size of evolved stars with confirmed planetary companions, refining the models of stellar-planetary interactions, and further investigating the mechanisms driving the ultimate fate of planets orbiting post-main-sequence stars.

\section*{Acknowledgments}

	This work is supported by National Key R\&D Progrom of China (grant No:2022YFE0116800),the National Natural Science Foundation of China (No. 11933008 and No. 12103084), the basic research project of Yunnan Province (Grant No. 202301AT070352). This paper includes data collected by the Kepler mission, which are publicly available from the Mikulski Archive for Space Telescopes (MAST)\footnote{https://mast.stsci.edu/}. 

   The data of MESA inlist is available on Zenodo under an open-source Creative Commons Attribution license: doi: 
   \href{https://doi.org/10.5281/zenodo.13941047}{https://doi.org/10.5281/zenodo.13941047}

\bibliography{sample631}{}
\bibliographystyle{aasjournal}

\newpage
\section{Appendix}\label{sec:Appendix}
	This is a supplementary materials to the paper, and it contains 2 tables (Table \ref{tab:exo_para},\ref{tab:nu0,nu2}) and 21 figures (Figure \ref{fig:exp_exoplanet}, \ref{fig:deltapi}, \ref{fig:pysyd_output}, \ref{fig:echelle}, \ref{fig:corner}).

   \renewcommand\thefigure{A}
   \begin{figure*}[ht!]
      \includegraphics[width=1.\linewidth]{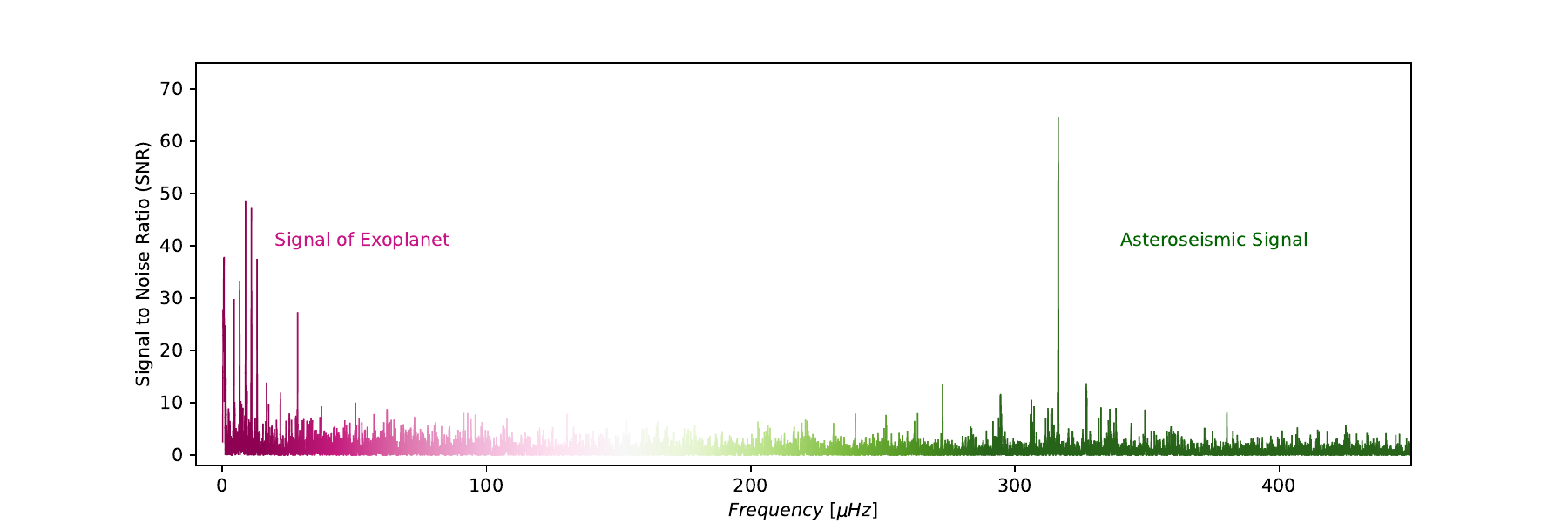}
      \centering
      \caption{The SNR diagram for Kepler-1004's light curve, shown as a demonstration, clearly displays the different signals. Since the exoplanet in this system was discovered via the transit method, the SNR diagram reveals the frequency components corresponding to the planet's orbital period and its harmonics, represented by the peaks in the pink part of the figure. The peaks in the green part display the solar-like oscillations of the red giant star. The harmonics of the planetary transit signal decay rapidly at higher frequencies and blend into the random noise, becoming invisible, thus not affecting the frequency analysis of the solar-like oscillations. The orbital periods of the planets in the other four systems studied in this work are even longer, resulting in lower signal frequencies, which are even less likely to interfere with the solar-like oscillations. To generate an SNR diagram that appropriately demonstrates both types of signals, we used a smoothing filter of width $log_{10}(1.6 \mu Hz)$ to compute the background noise.\label{fig:exp_exoplanet}}
   \end{figure*}

   \renewcommand\thetable{A}
   \begin{deluxetable*}{lccc}[ht!]
      \setlength{\tabcolsep}{8pt}
      \tablecaption{Orbital information of exoplanets and candidates\label{tab:exo_para}}
      \tablehead{
        \colhead{Exoplanet Name} & \colhead{Source} &\colhead{P} & \colhead{e}  \\
        \colhead{} & \colhead{} & \colhead{(day)} & \colhead{}
      }
      \startdata
      KOI-75.01     & Q1-Q17 DR25 KOI Table & $105.88177\pm0.00013$ & 0  \\
      Kepler-643 b  & \citep{2018ApJ...861L...5G} & $16.338896\pm0.000018$ & $0.37\pm0.06$\\
      Kepler-815 b  & \citep{2019RAA....19...41G} & $8.574386\pm0.000044$ & \nodata \\
      Kepler-1004 b   & \citep{2019RAA....19...41G} & 	$5.287854\pm0.000013$ & \nodata \\
      KOI-2640  & Q1-Q8 KOI Table & $33.1809\pm0.0014$ & 0 \\
      \enddata
      \tablecomments{This is the \href{https://exoplanetarchive.ipac.caltech.edu/docs/Kepler_KOI_docs.html}{website} of Q1-Q17 DR25 KOI Table and Q1-Q8 KOI Table}
   \end{deluxetable*}

   \renewcommand\thetable{B}
   \begin{longrotatetable}
      \linespread{1}
      \begin{deluxetable*}{lccccccccc}
      \tablecaption{$\nu_{n,0}$ and $\nu_{n,2}$ of Host Stars identified by \texttt{PBjam} \label{tab:nu0,nu2}}
      \tablehead{
      \colhead{Host Name} & \colhead{$l$} & \colhead{1 ($\mu Hz$)} & \colhead{2 ($\mu Hz$)} & \colhead{3 ($\mu Hz$)} & \colhead{4 ($\mu Hz$)} & \colhead{5 ($\mu Hz$)} & \colhead{6 ($\mu Hz$)} & \colhead{7 ($\mu Hz$)} & \colhead{8 ($\mu Hz$)}
      }
      \startdata
      KOI-75 & 0 & $477.015 \pm 0.183$ & $519.515 \pm 0.112$ & $552.746 \pm 0.080$ & $591.261 \pm 0.064$ & $629.685 \pm 0.056$ & $668.861 \pm 0.081$ & $707.579 \pm 0.156$ & $747.227 \pm 0.260$ \\
         & 2 & $472.067 \pm 0.899$ & $514.043 \pm 0.645$ & $549.524 \pm 0.195$ & $587.954 \pm 0.280$ & $625.784 \pm 0.186$ & $665.510 \pm 0.200$ & $698.287 \pm 0.209$ & $744.08 \pm 0.417$ \\
      Kepler-643 & 0 & $418.029 \pm 0.883$ & $450.163 \pm 0.264$ & $483.927 \pm 0.065$ & $517.184 \pm 0.021$ & $550.524 \pm 0.037$ & $584.209 \pm 0.041$ \\
         & 2 &  $414.449 \pm 0.998$ & $446.656 \pm 0.490$ & $481.284 \pm 0.286$ & $514.493 \pm 0.052$ & $546.545 \pm 0.047$ & $580.872 \pm 0.068$\\
      Kepler-815 & 0 & $244.886 \pm 0.683$ & $268.188 \pm 0.478$ & $290.948 \pm 0.042$ & $314.218 \pm 0.030$ & $337.554 \pm 0.018$ & $360.837 \pm 0.016$ & $384.355 \pm 0.030$ & $408.142 \pm 0.286$ \\
         & 2 & $242.510 \pm 0.650$ & $265.478 \pm 0.691$ & $288.214 \pm 0.194$ & $311.752 \pm 0.041$ & $335.092 \pm 0.035$ & $358.393 \pm 0.178$ & $382.336 \pm 0.144$ & $405.921 \pm 0.115$ \\
      Kepler-1004 & 0 & $229.361 \pm 0.666$ & $251.195 \pm 0.250$ & $272.626 \pm 0.025$ & $294.582 \pm 0.075$ & $316.368 \pm 0.028$ & $338.257 \pm 0.065$ \\
         & 2 & $226.740 \pm 0.654$ & $248.408 \pm 0.530$ & $269.988 \pm 0.388$ & $291.693 \pm 0.036$ & $312.523 \pm 0.282$ & $336.016 \pm 0.152$\\
      KOI-2640 & 0 & $53.881 \pm 0.030$ & $61.008 \pm 0.017$ & $68.278 \pm 0.012$ & $75.780 \pm 0.010$ & $83.175 \pm 0.015$ & $90.805 \pm 0.022$ & $98.463 \pm 0.083$ \\
         & 2 &  $52.625 \pm 0.133$ & $59.974 \pm 0.030$ & $67.223 \pm 0.018$ & $74.741 \pm 0.018$ & $82.232 \pm 0.023$ & $89.873 \pm 0.034$ & $97.490 \pm 0.089$\\
      \enddata
      \end{deluxetable*}
      \tablecomments{The column numbers (1, 2, 3, …) in this table are for sequential ordering purposes only and do not correspond to the radial order $n$.}
   \end{longrotatetable}

   \renewcommand\thefigure{B}
   \begin{figure*}[ht!]
    \gridline{\fig{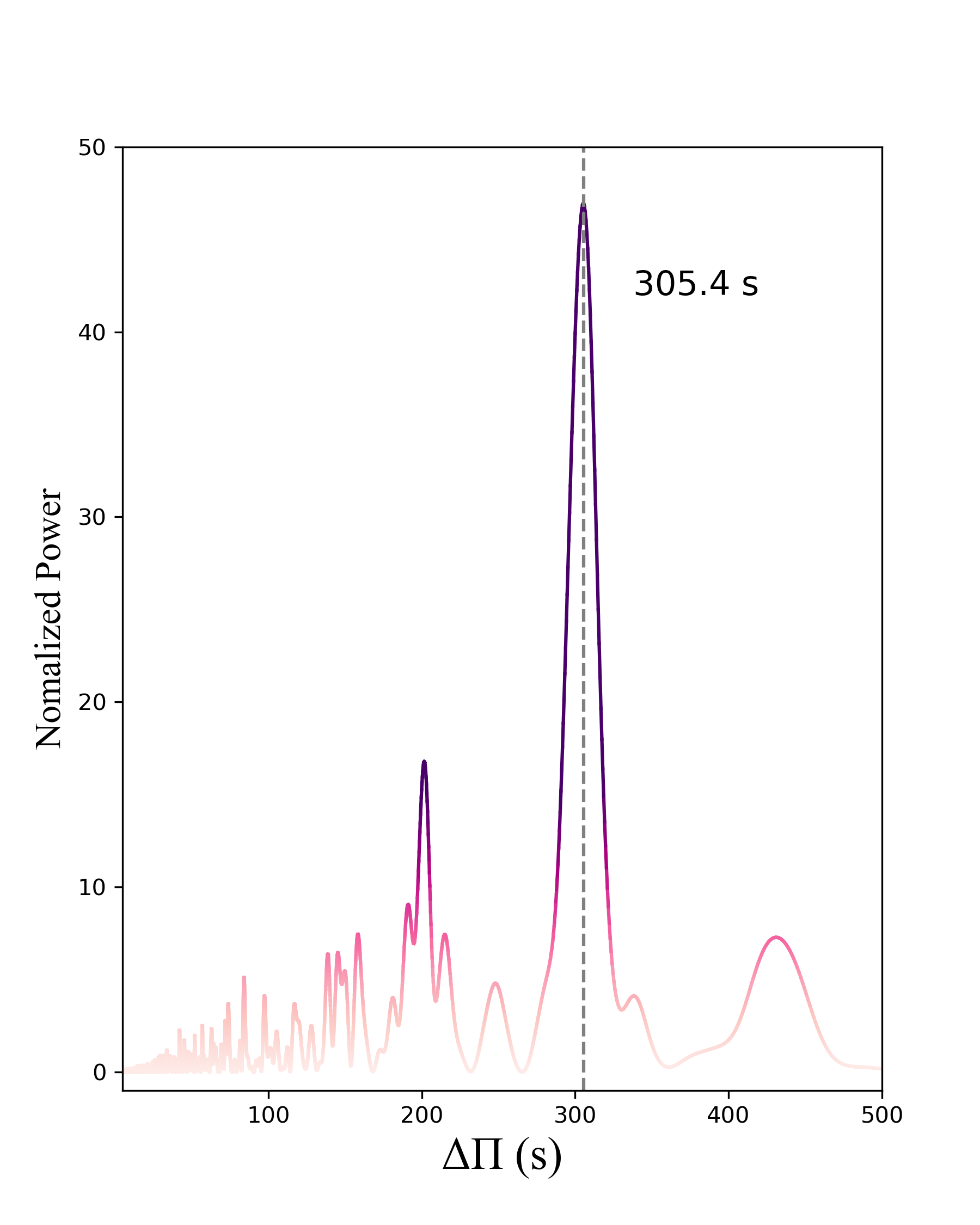}{0.33\textwidth}{KOI-75}
    \fig{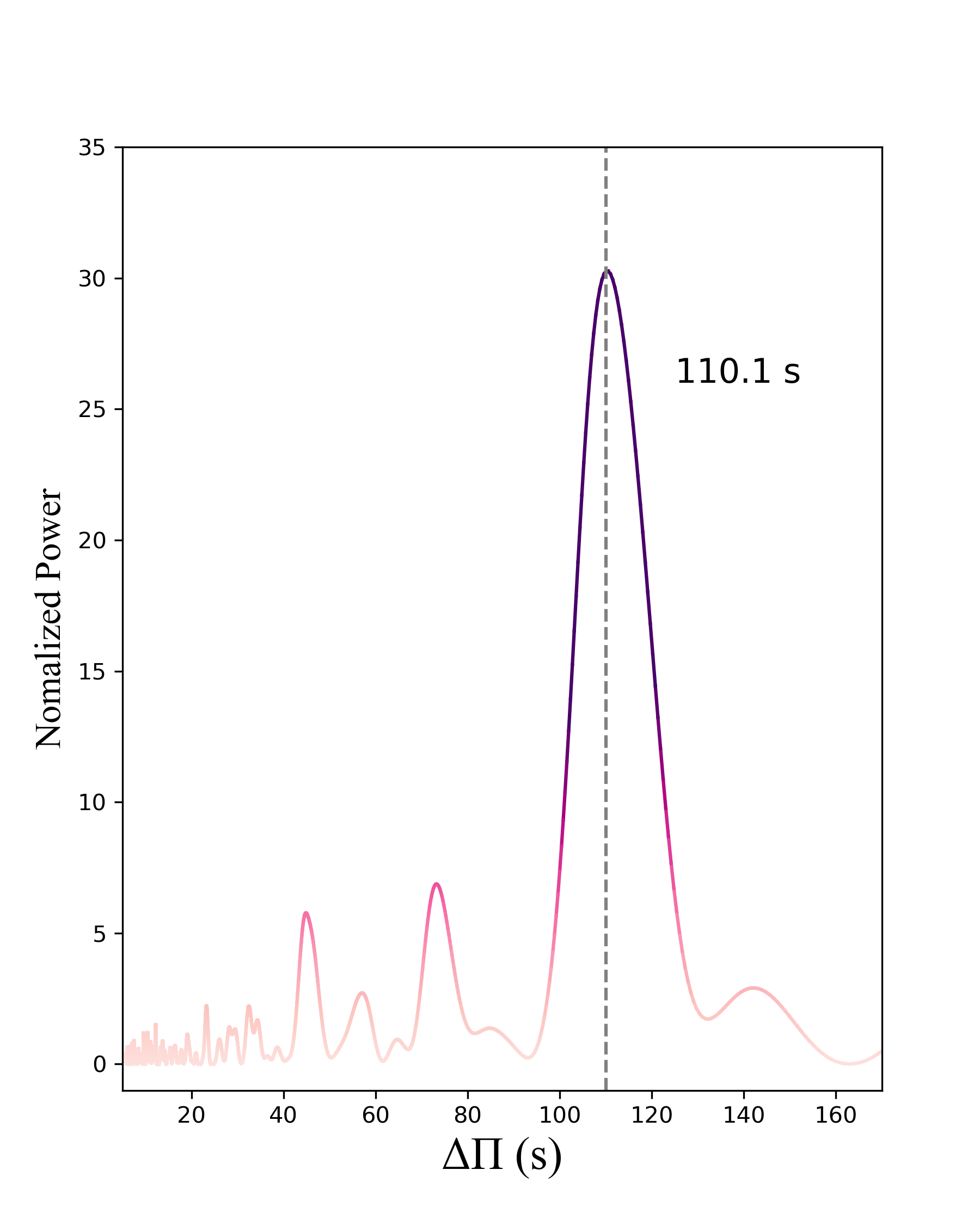}{0.33\textwidth}{Kepler-643}
    \fig{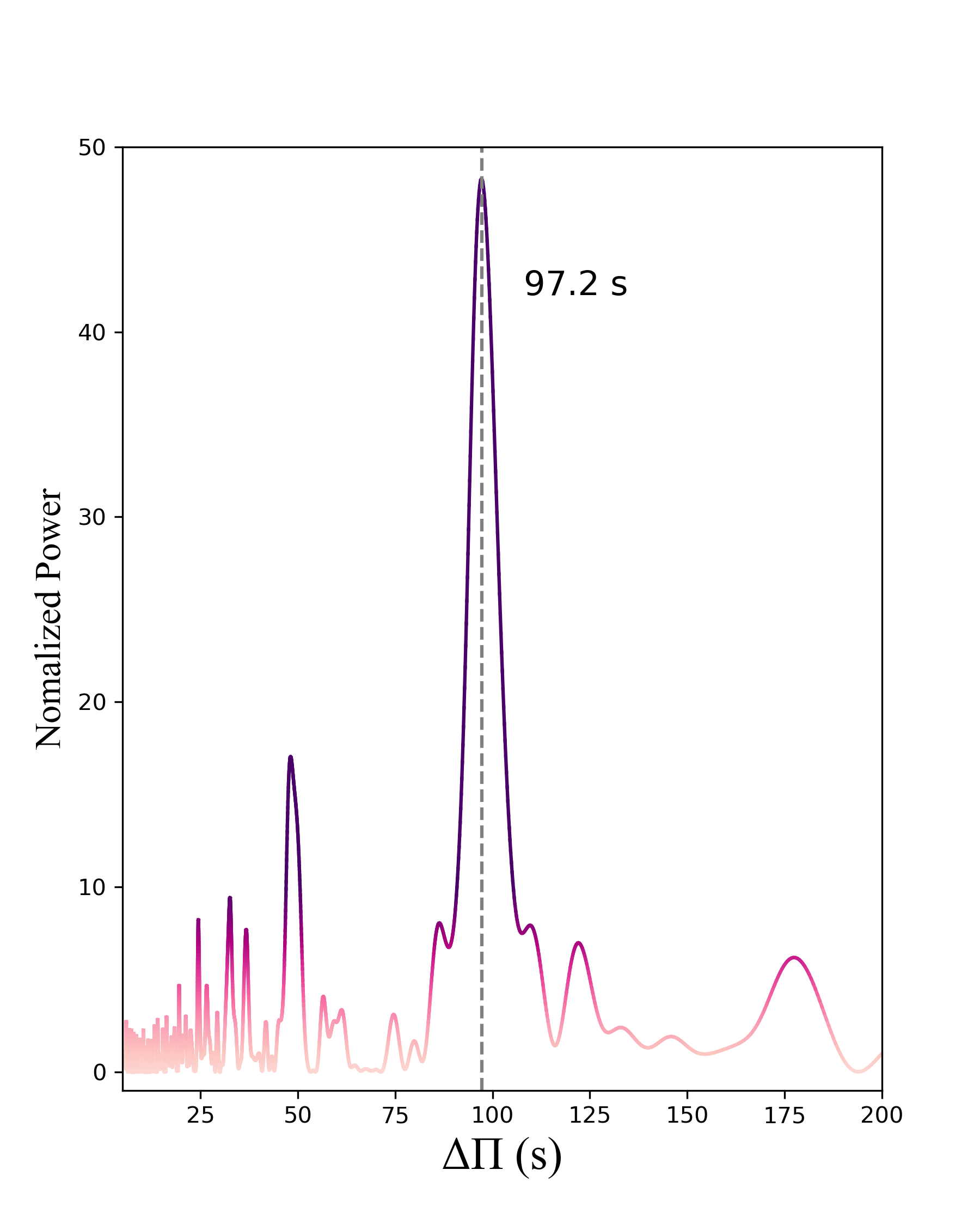}{0.33\textwidth}{Kepler-815}
    }
    \gridline{\fig{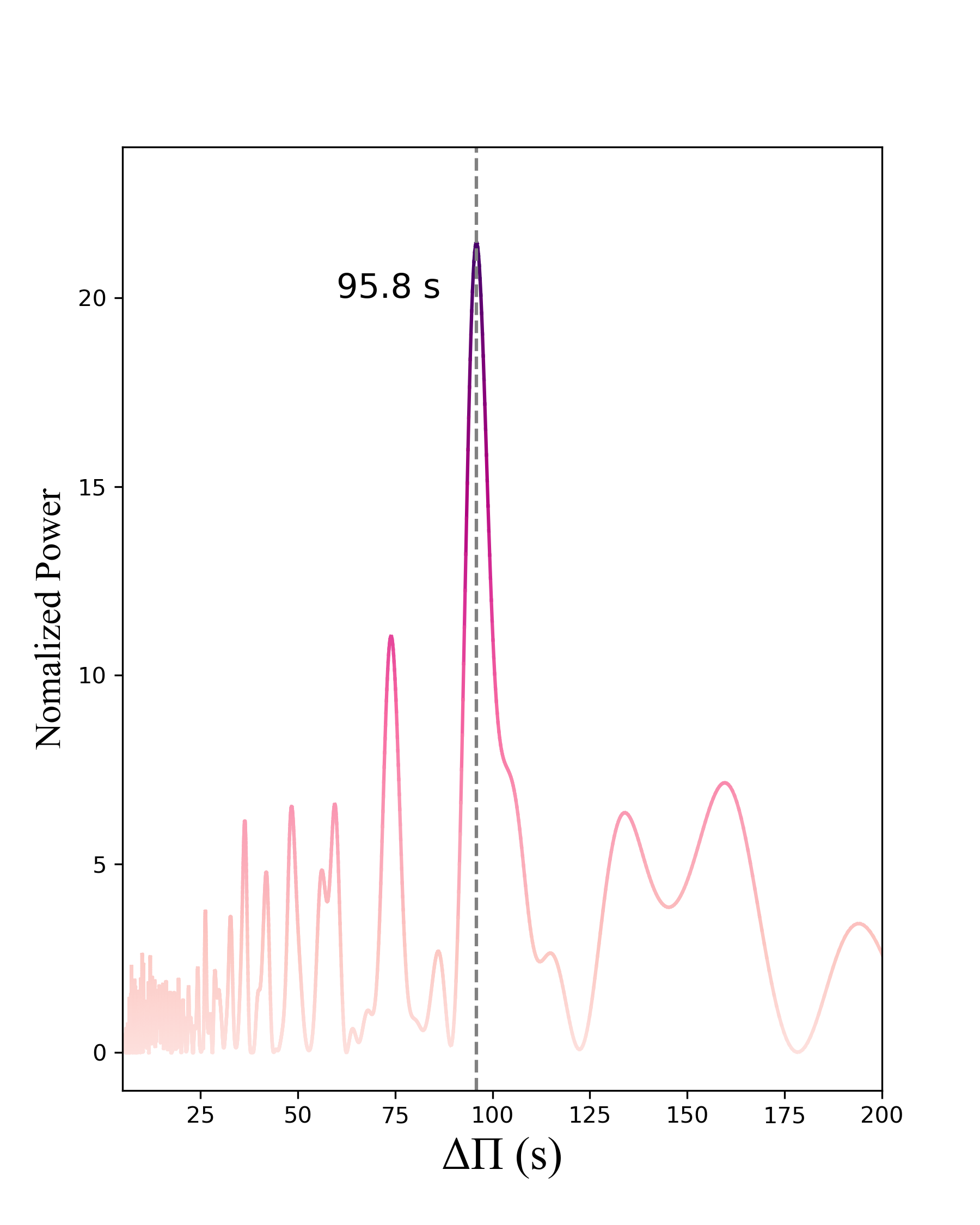}{0.33\textwidth}{Kepler-1004}
    \fig{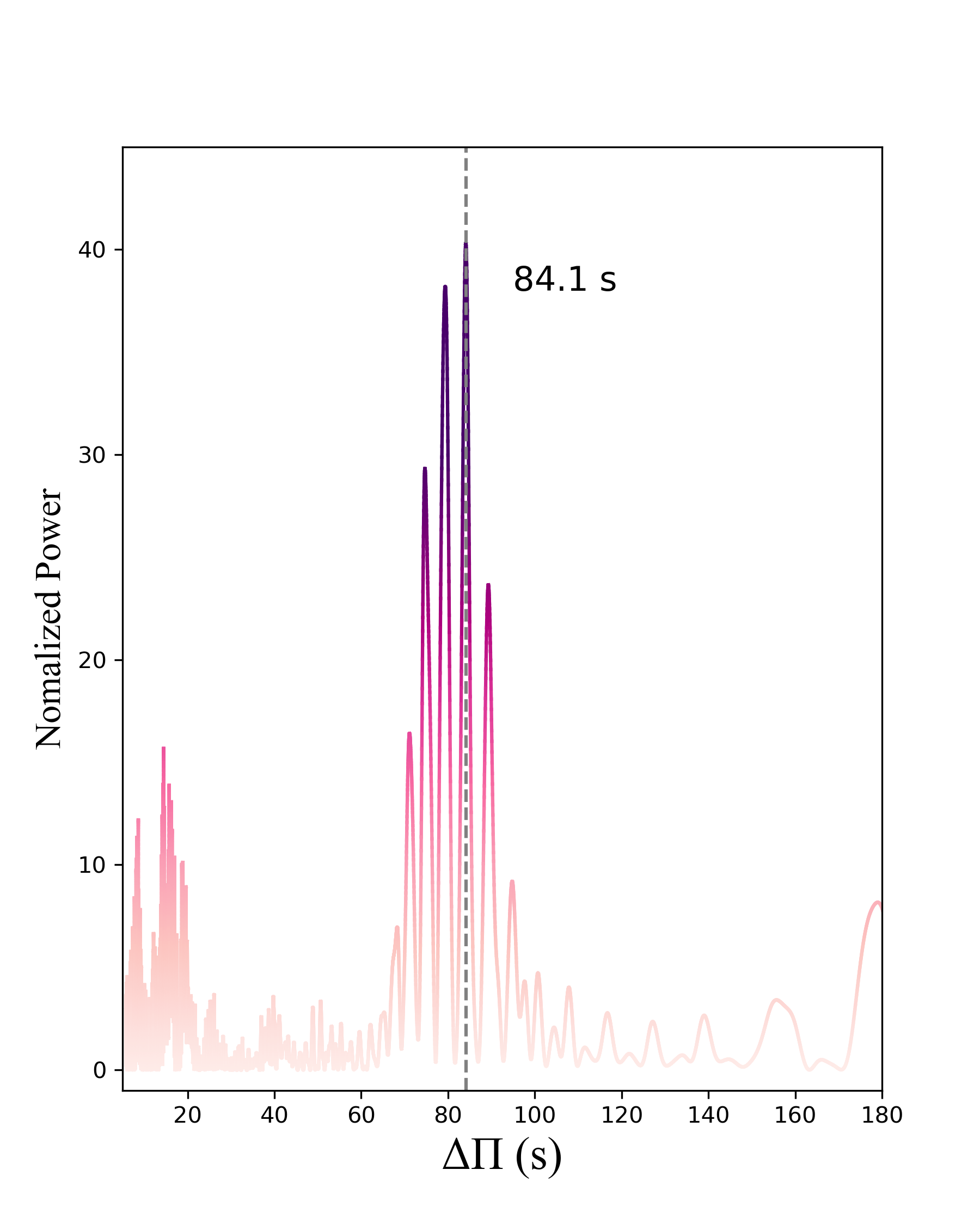}{0.33\textwidth}{KOI-2640}
    }
    \caption{The power spectrum of $P(\tau)$  for five host stars\label{fig:deltapi}}
   \end{figure*}

   \renewcommand\thefigure{C}
   \begin{figure*}[ht!]
    \gridline{\fig{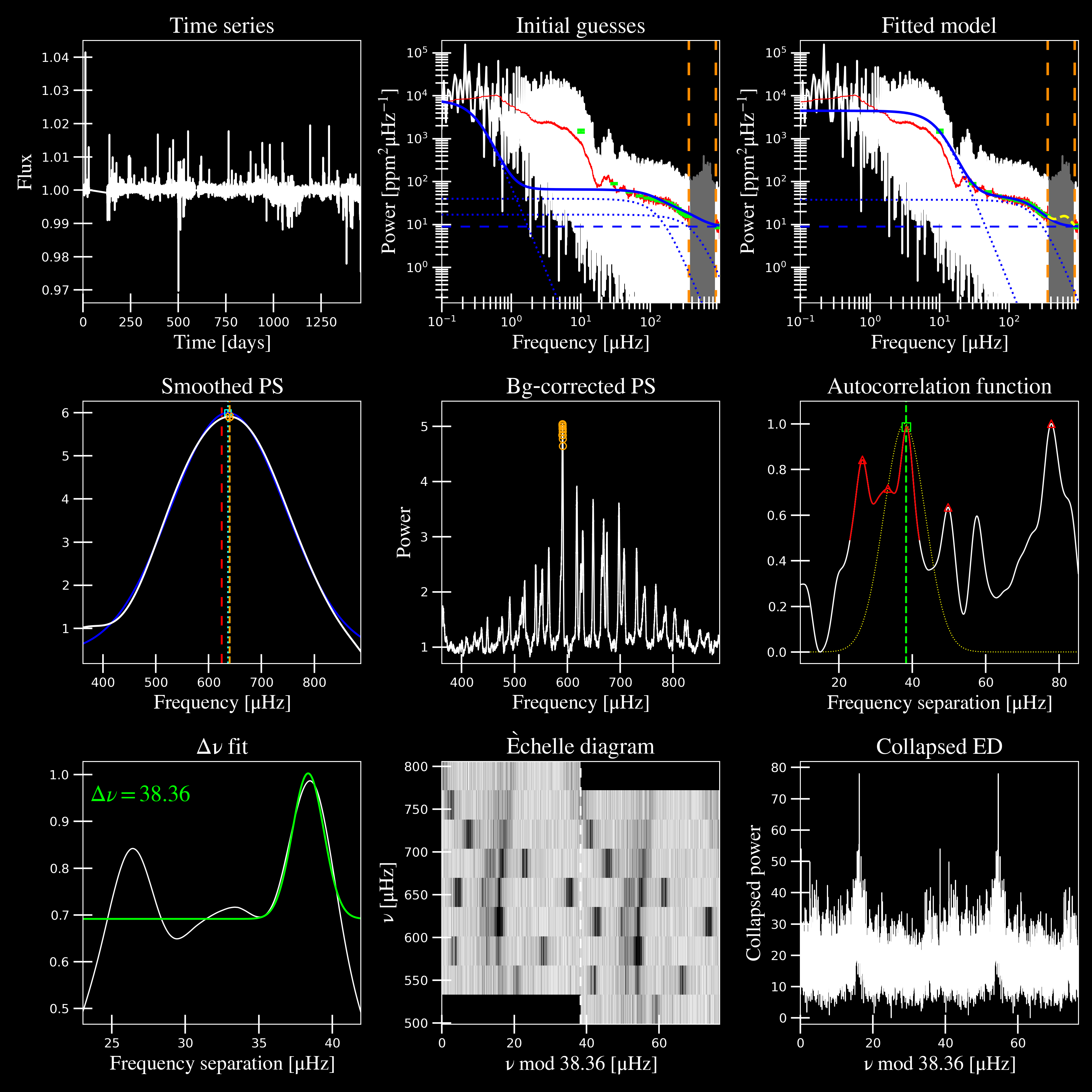}{0.38\textwidth}{KOI-75}
    \fig{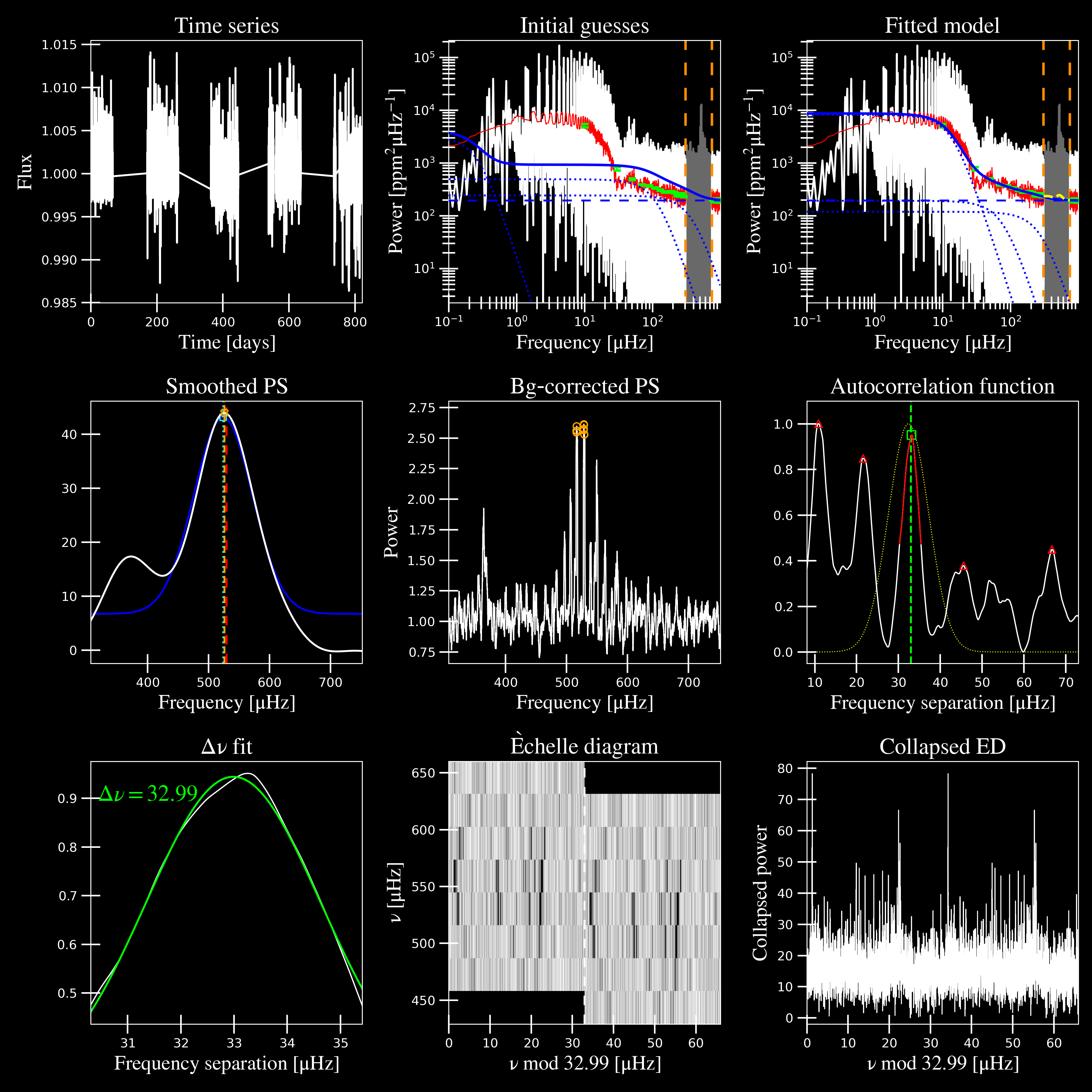}{0.38\textwidth}{Kepler-643}
    }
    \gridline{\fig{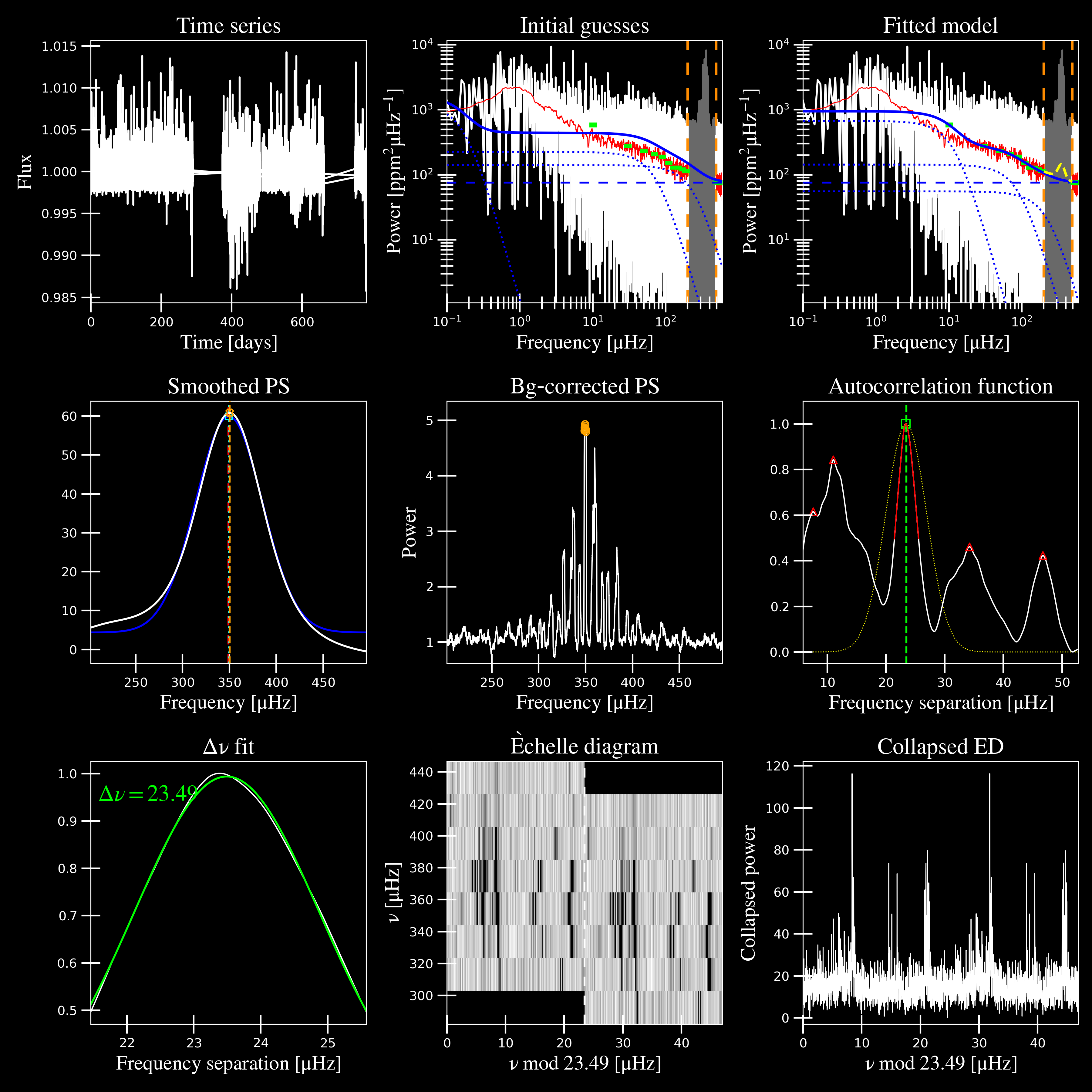}{0.38\textwidth}{Kepler-815}
    \fig{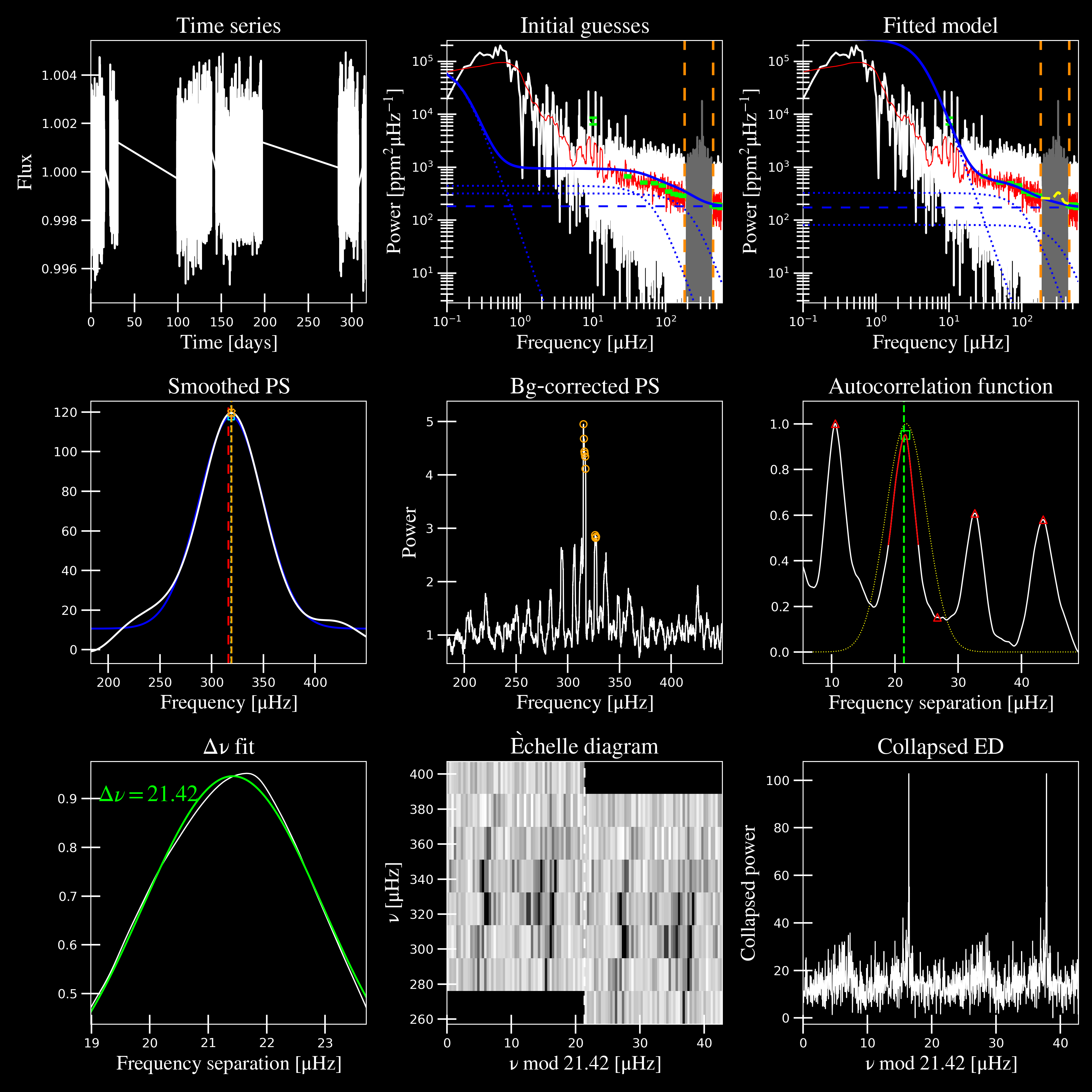}{0.38\textwidth}{Kepler-1004}
    }
    \gridline{\fig{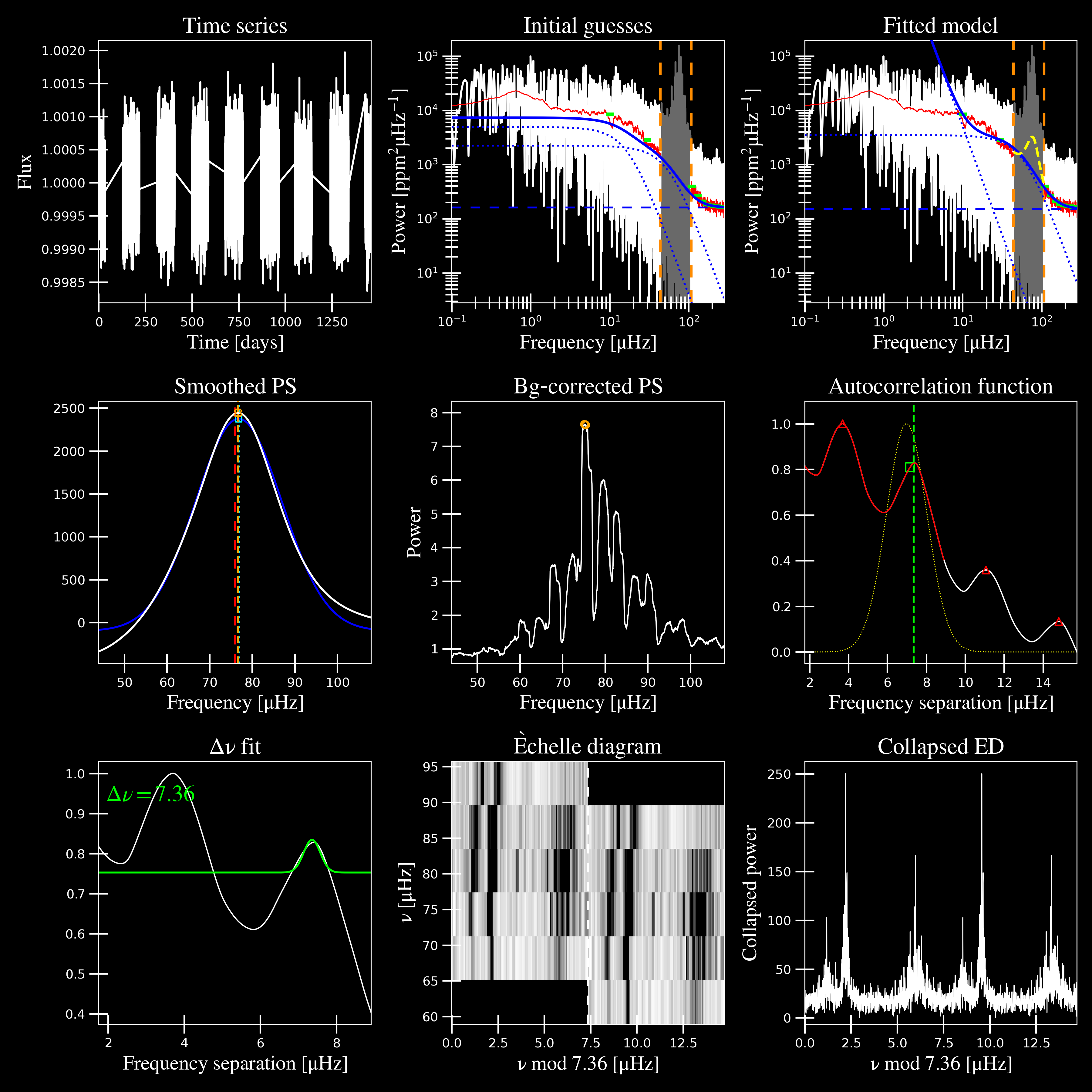}{0.38\textwidth}{KOI-2640}
    }
    \caption{Outputs of the asteroseismic parameters obtained after analyzing the power spectrum with the \texttt{pySYD} package, Detailed descriptions of each figure can be found at this \href{https://pysyd.readthedocs.io/en/latest/library/output.html}{website}.\label{fig:pysyd_output}}
   \end{figure*}

   \renewcommand\thefigure{D}
   \begin{figure*}[ht!]
    \gridline{\fig{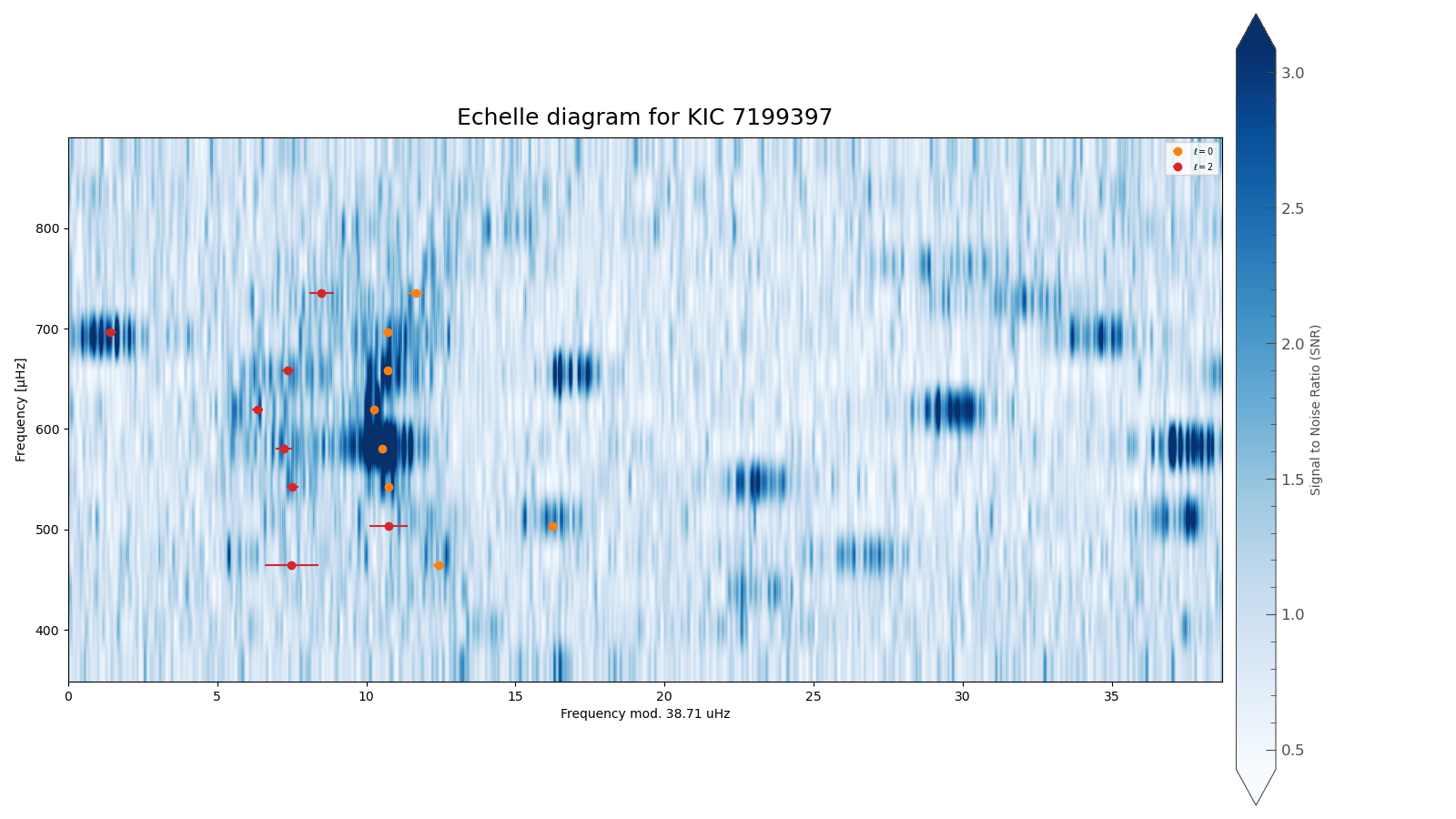}{0.49\textwidth}{KOI-75}
    \fig{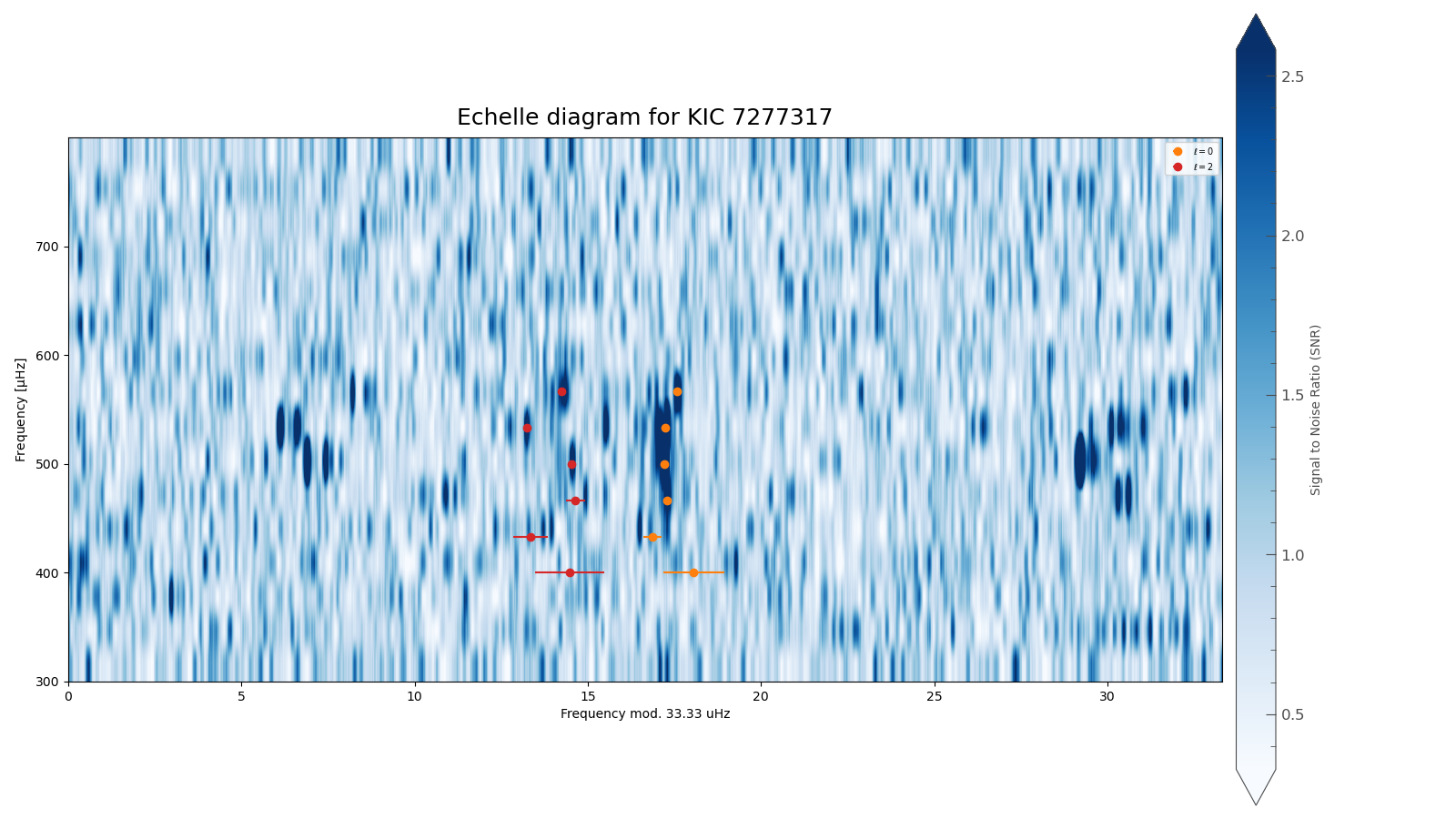}{0.49\textwidth}{Kepler-643}
    }
    \gridline{\fig{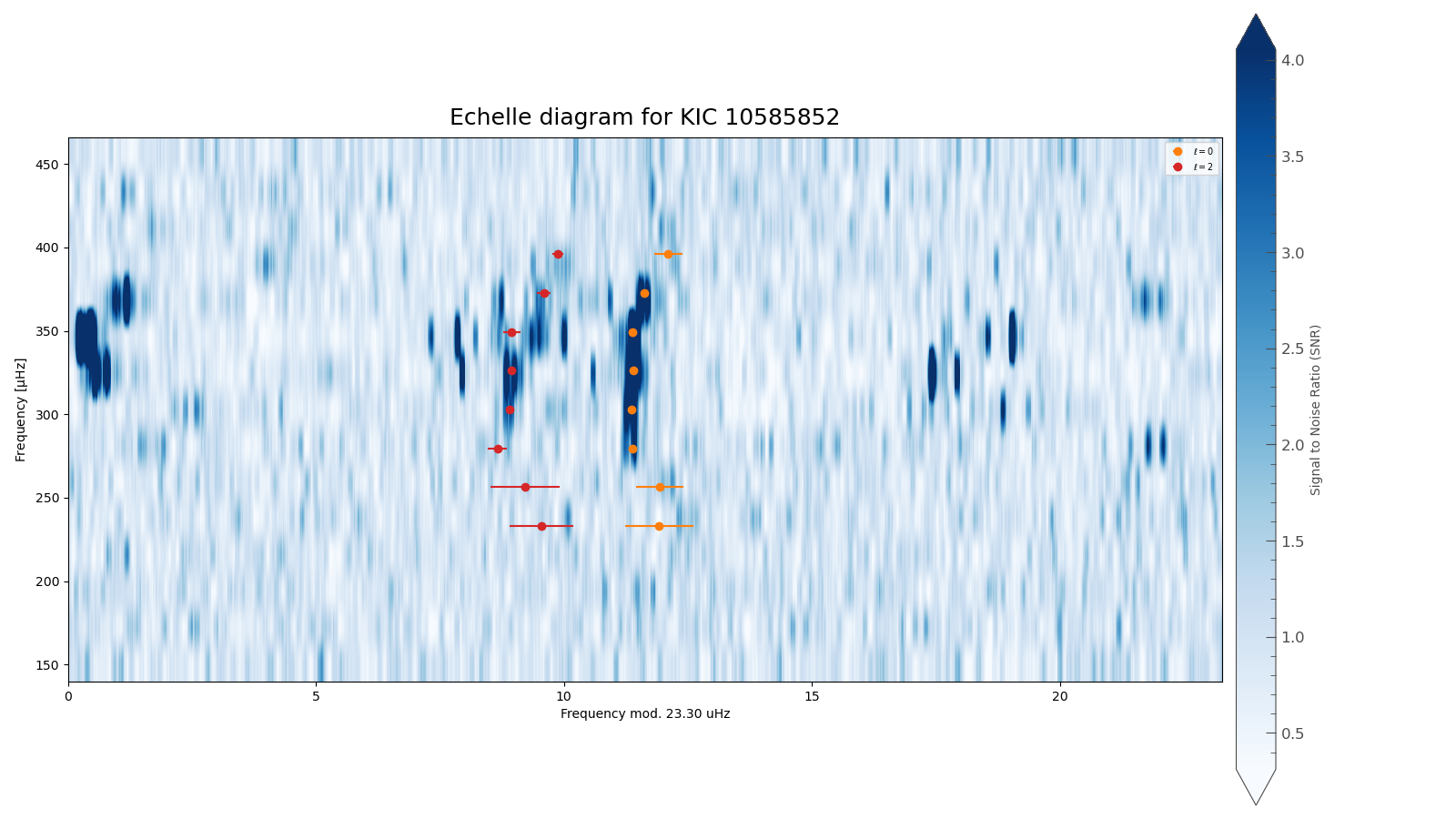}{0.49\textwidth}{Kepler-815}
    \fig{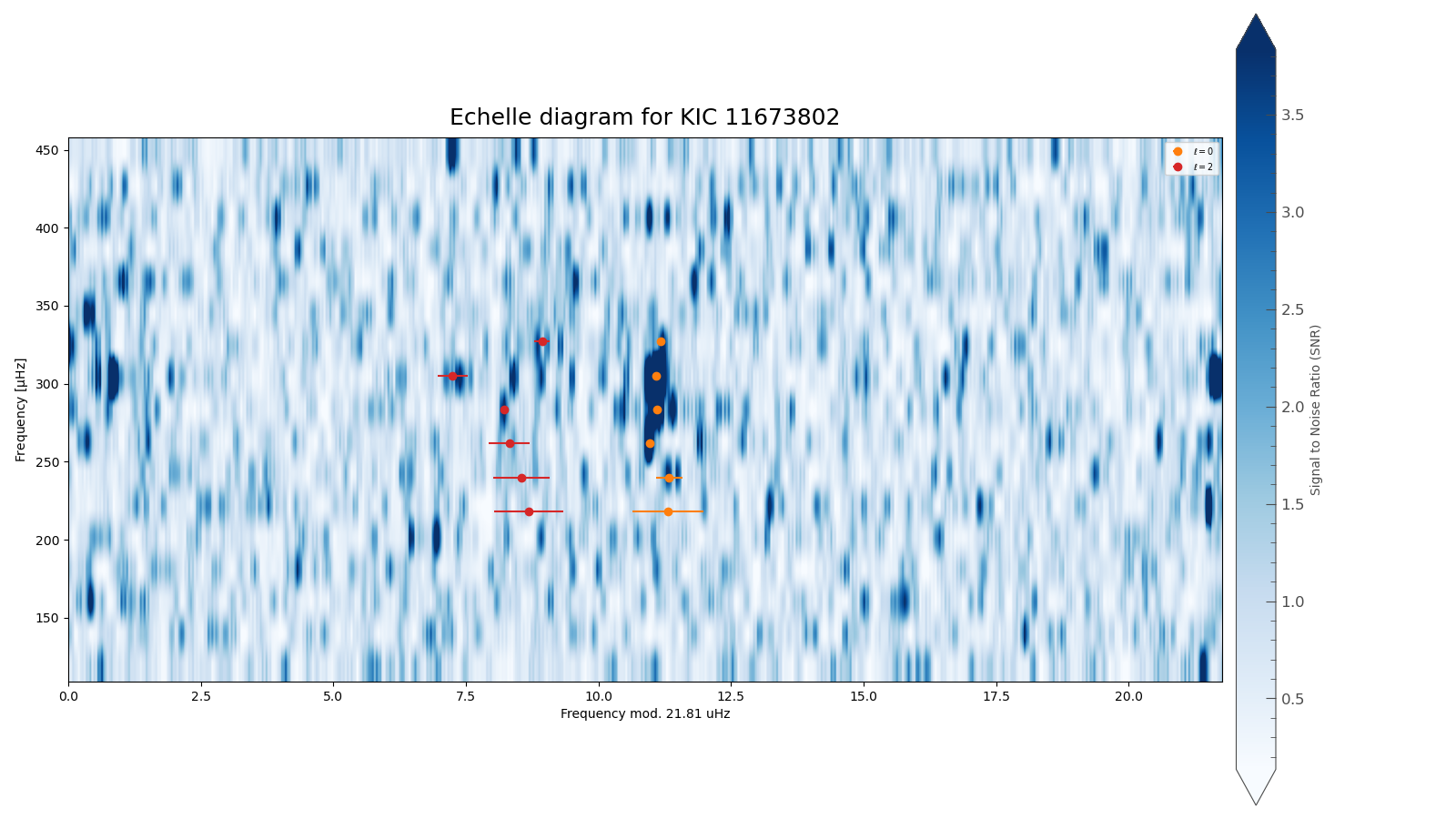}{0.49\textwidth}{Kepler-1004}
    }
    \gridline{\fig{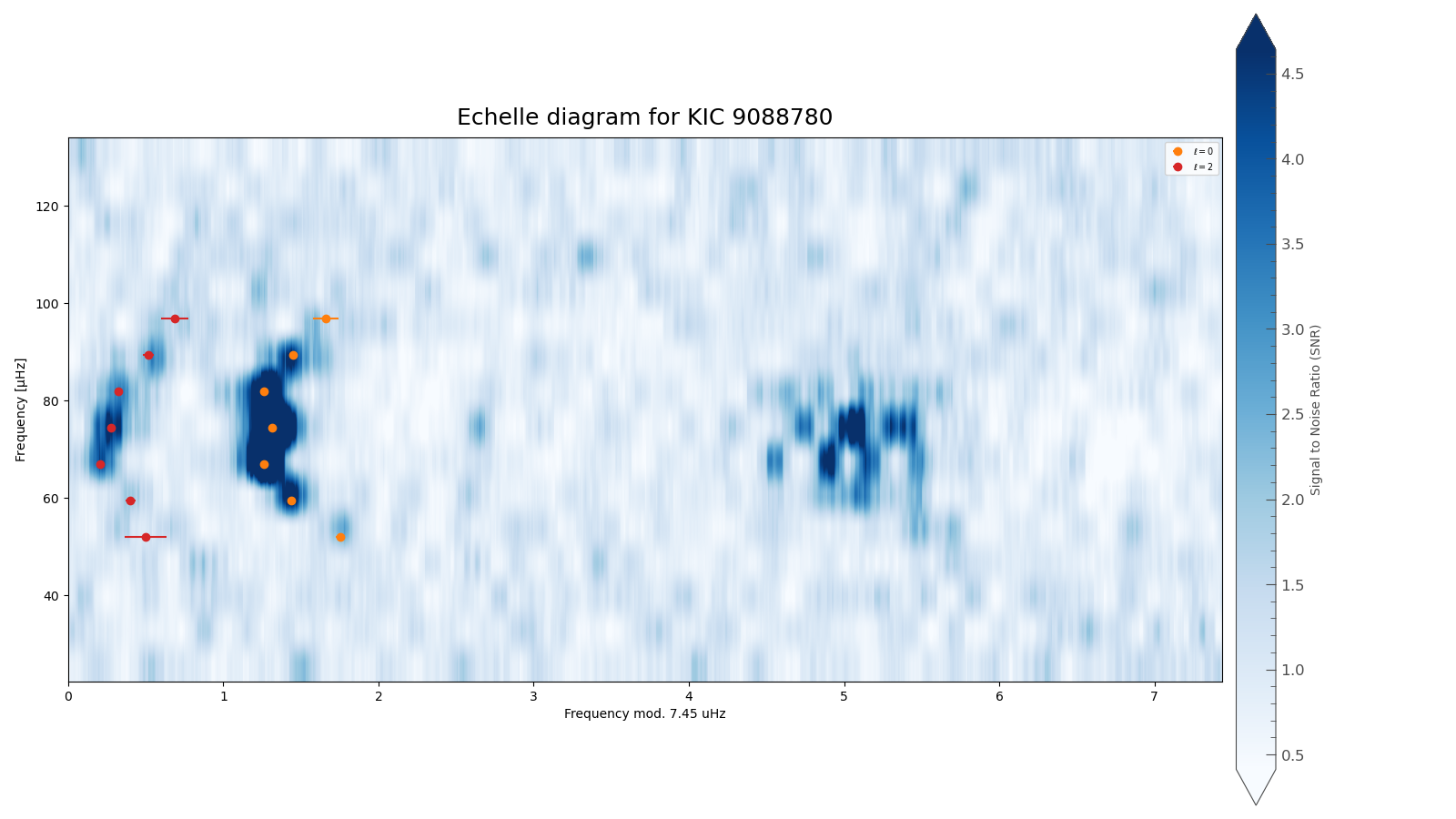}{0.49\textwidth}{KOI-2640}
    }
    \caption{The echelle diagrams of host stars, the orange and red dots represent $\nu_{n,0}$ and $\nu_{n,2}$ identified by \texttt{PBjam}.\label{fig:echelle}}
   \end{figure*}

   \renewcommand\thefigure{E}
   \begin{figure*}[ht!]
    \gridline{\fig{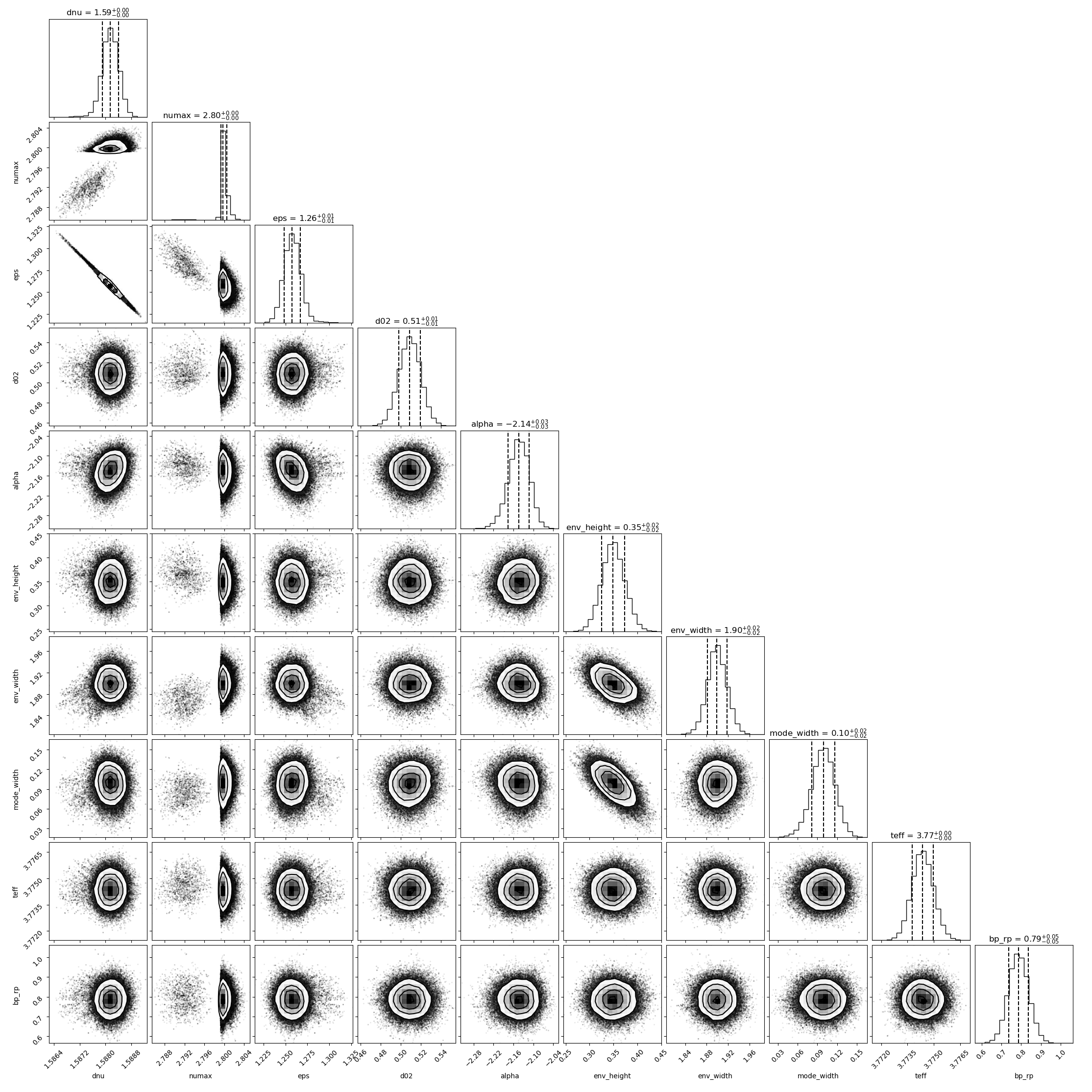}{0.4\textwidth}{KOI-75}
    \fig{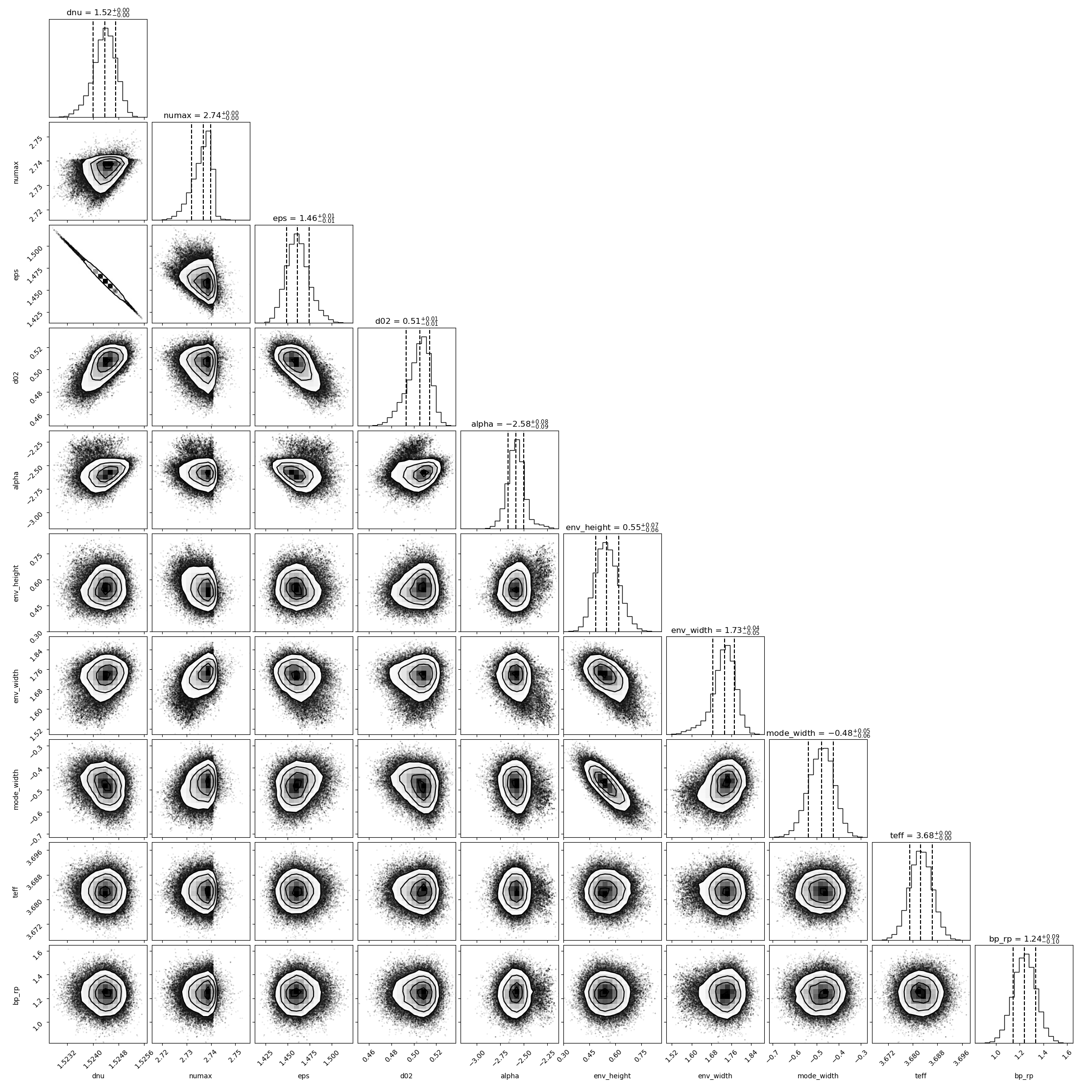}{0.4\textwidth}{Kepler-643}
    }
    \gridline{\fig{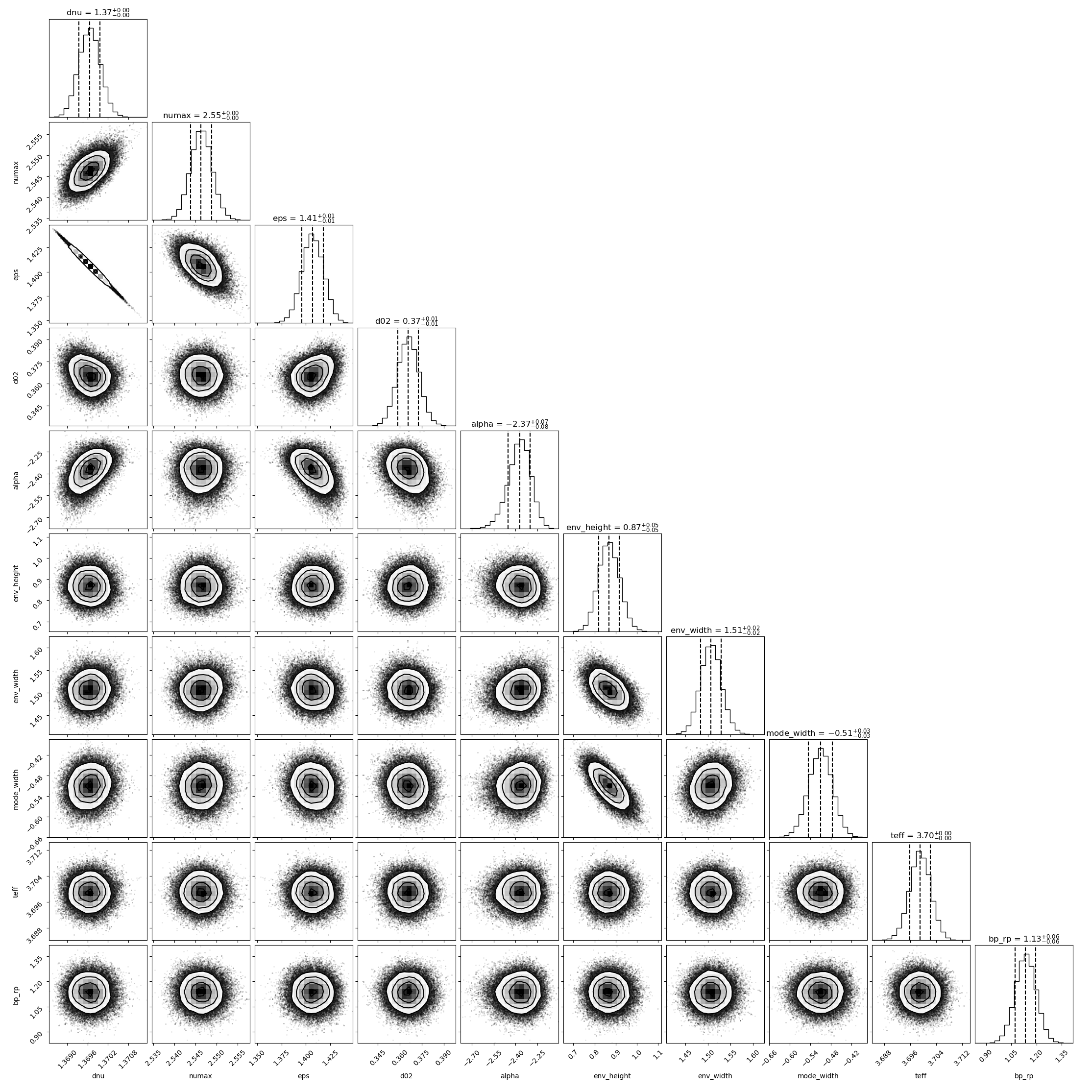}{0.4\textwidth}{Kepler-815}
    \fig{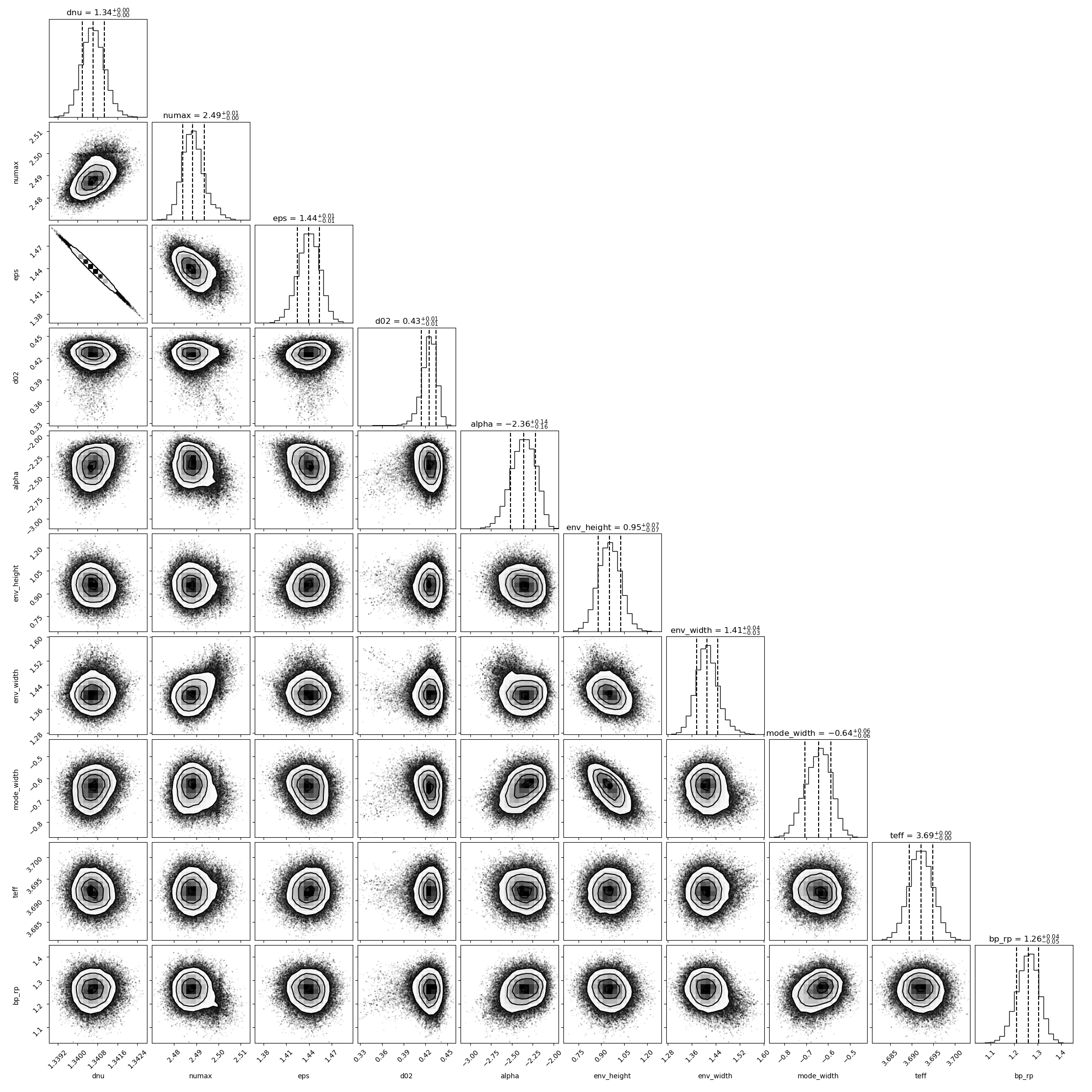}{0.4\textwidth}{Kepler-1004}
    }
    \gridline{\fig{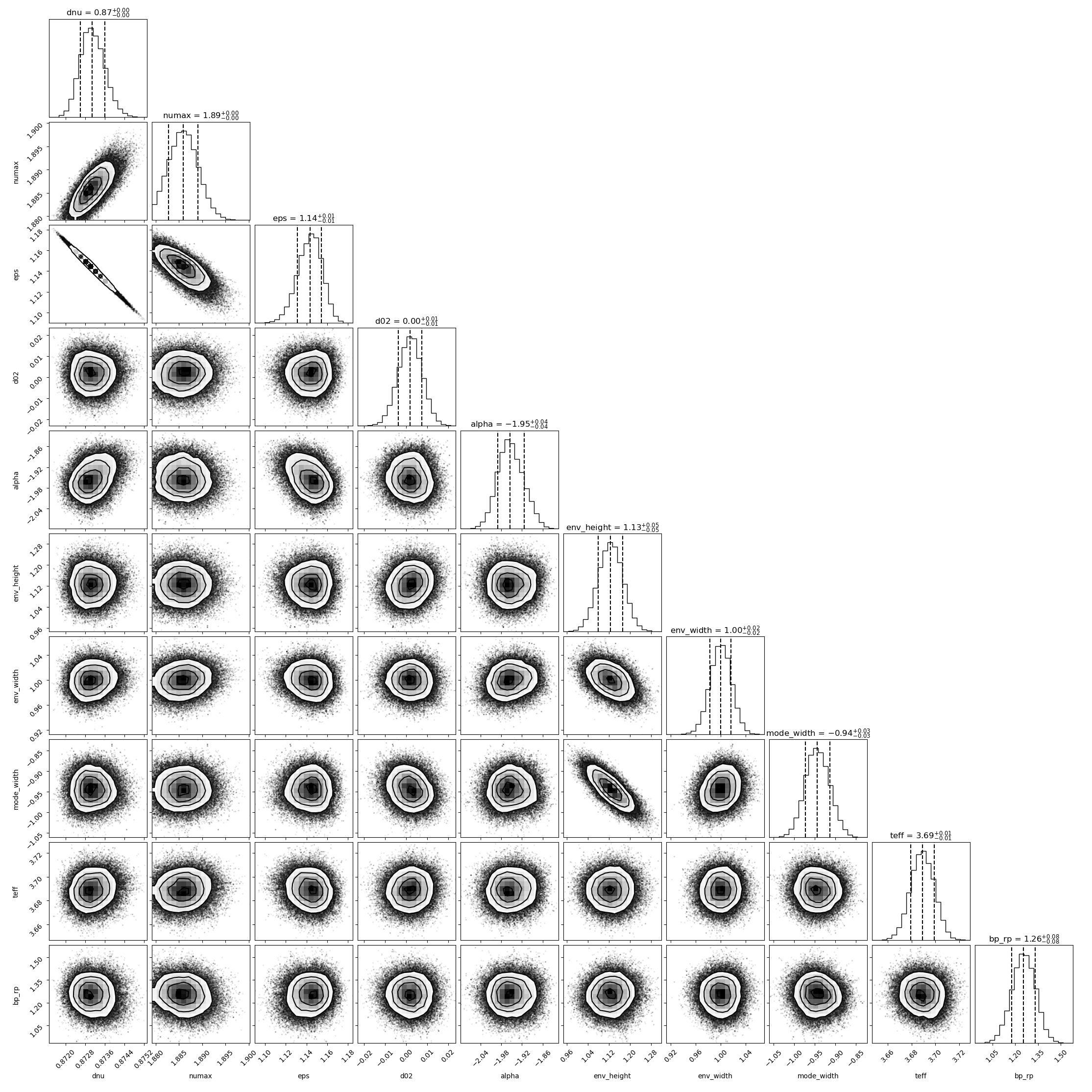}{0.4\textwidth}{KOI-2640}
    }
    \caption{Corner diagram of asymptotic fit parameters\label{fig:corner}}
   \end{figure*}

\end{document}